# Geoscience for understanding habitability in the solar system and beyond

Veronique Dehant[1, 2], Vinciane Debaille[3], Vera Dobos[4, 5, 6], Fabrice Gaillard[7], Cedric Gillmann[1, 3], Steven Goderis[8], John Lee Grenfell[9], Dennis Höning[9, 10], Emmanuelle J. Javaux[11], Özgür Karatekin[1], Alessandro Morbidelli[12], Lena Noack[1, 13], Heike Rauer[9, 13, 14], Manuel Scherf[13], Tilman Spohn[9], Paul Tackley[16], Tim Van Hoolst[1], Kai Wünnemann[13, 17]

[1] Royal Observatory of Belgium
[2] Université Catholique de Louvain, Belgium
[3] Université Libre de Bruxelles, Belgium
[4] Konkoly Thege Miklós Astronomical Institute, Research Centre for Astronomy and Earth Sciences, Hungarian Academy of Sciences, Budapest, Hungary
[5] Geodetic and Geophysical Institute, Research Centre for Astronomy and Earth Sciences, Hungarian Academy of Sciences, Sopron, Hungary
[6] MTA-ELTE Exoplanet Research Group, 9700, Szent Imre h. u. 112, Szombathely, Hungary
[7] University of Orléans, France
[8] Vrije Universiteit Brussel, Belgium
[9] German Aerospace Centre (DLR), Germany
[10] Vrije Universiteit Amsterdam
[11] University of Liège, Belgium
[12] University of Nice, France
[13] Free University of Berlin, Germany
[14] Berlin Institute of Technology, Germany
[15] Austrian Academy of Sciences, Austria
[16] ETH Zürich, Switzerland
[17] Museum of Natural History, Germany

## Contents

## Abstract

This paper reviews habitability conditions for a terrestrial planet from the point of view of geosciences. It addresses how interactions between the interior of a planet or a moon and its atmosphere and surface (including hydrosphere and biosphere) can affect habitability of the celestial body. It does not consider in detail the role of the central star but focusses more on surface conditions capable of sustaining life. We deal with fundamental issues of planetary habitability, i.e. the environmental conditions capable of sustaining life, and the above-mentioned interactions can affect the habitability of the celestial body.

We address some hotly debated questions including:

- How do core and mantle affect the evolution and habitability of planets?
- What are the consequences of mantle overturn on the evolution of the interior and atmosphere?
- What is the role of the global carbon and water cycles?
- What influence do comet and asteroid impacts exert on the evolution of the planet?
- How does life interact with the evolution of the Earth's geosphere and atmosphere?

- How can knowledge of the solar system geophysics and habitability be applied to exoplanets?

In addition, we address the identification of preserved life tracers in the context of the interaction of life with planetary evolution.

## Introduction

The bulk evolution of a planet (including the Earth) is driven by the internal energy sources and losses (radiogenic sources, energy from accretion and differentiation, tidal energy), and depends upon the composition, structure, and thermal state of its core, mantle, lithosphere, crust. These properties depend on the size and mass of the planet as well as on potential interactions with a possible ocean and atmosphere and – in the case of the Earth or other currently unknown planets possibly having life – with a biosphere. In addition, impacts (in particular, giant collision events) may have had a sustainable effect on the evolution of lithospheres, atmospheres, and biospheres.

The paper has a different focus than most other habitability discussions in Earth-System sciences, planetary sciences and astronomy by encompassing the entire planet from the upper atmosphere to the deep interior and by including the exogenic influence within the framework of the study of its habitability. In this aspect, it is broader in scope than e.g., Lammer et al. (2009), Javaux and Dehant (2010), Spohn (2014), Fritz et al. (2014), Southam et al. (2015), Cockell et al. (2016), and Tosi et al. (2017). The paper addresses questions within six main sections: (1) Formation of habitable planets, (2) role of cometary, meteorite and asteroid impacts on planetary evolution, (3) planetary interiors, mantle convection and their evolutions, (4) atmospheric evolution, (5) interaction of life with the atmosphere, geosphere and interior of planets, (6) identification of preserved life tracers in the context of the interaction of life with planetary evolution, (7) conclusion and implications for planet formation and habitability conditions and detection in exoplanetary systems.

## 1. Formation of habitable planets

The Solar System and planetary systems around other stars are built from an initial protoplanetary disk around the central star containing gas and dust. Dust grows to pebbles by coagulation and deposition of volatile ices, but the continued growth to planetesimals is, for instance, hampered by the poor sticking of mm-to-cm-sized pebbles. Planetesimals can nevertheless form by gravitational collapse of pebble clumps concentrated in the turbulent gas through the streaming instability (Yang et al., 2017). The subsequent growth initially occurs by planetesimal-planetesimal collisions, but it is the accretion rate of pebbles which dominates the growth of larger (~1000-km-sized) protoplanets (for pebble accretion, see Ormel and Klahr, 2010; Lambrechts and Johansen, 2012) to form terrestrial planetary embryos and the solid cores of gas giants, ice giants, and super-Earths (extrasolar planets with a mass up to 10 times the Earth's, but by an order of magnitude smaller than the masses of the Solar System's ice giants like Jupiter.

The formation of the terrestrial planets could begin with planetesimals distributed throughout the inner Solar System and the asteroid belt. However, in this case it would end with 0.5–1.0 Earth-masses planets inside 2 AU (Astronomical Unit), much larger than the actual mass of Mars, namely 0.1 Earth-masses (Raymond et al., 2009). Obtaining the order of magnitude mass-contrast between the Earth and Mars requires that the planetesimals were concentrated within 1 AU from the Sun, with little total mass beyond this limit (Hansen, 2009). This planetesimal concentration can be obtained in two ways. The "grand tack hypothesis" (Walsh et al., 2011; Pierens et al., 2014) proposes that, after its formation at ~3.5 AU, Jupiter migrated inward to 1.5 AU, before reversing course due to capturing Saturn in an orbital resonance (Masset and Snellgrove, 2001; Morbidelli and Crida, 2007), eventually halting near its current orbit at 5.2 AU. Due to Jupiter's migration, the inner planetesimal disk was truncated at 1.0 AU, leaving behind only a mass-depleted and dynamically excited asteroid belt beyond this radius. The second possibility is that the streaming instability was effective only inwards of 1 AU (Drazkowska et al., 2016). Beyond this limit, asteroids could form only later, at the time of the photo-evaporation of the disk (Carrera et al., 2017), producing a low-mass population of objects. The dynamical excitation of the asteroid belt in this case requires some chaotic evolution of Jupiter for a period lasting a few My (Izidoro et al., 2016). Which of these two scenarios is more appropriate is the subject of ongoing debate.

Our planet formed slowly over tens of millions of years, as indicated by radioactive chronometers (Kleine et al., 2009) and explained by models of collisional accretion of a disk of planetesimals and planetary embryos (e.g., O'Brien et al., 2006; Jacobson et al., 2014). The Earth formed mostly after the disappearance of the protoplanetary disk of gas, via a sequence of giant impacts. The precursors of the Earth, the planetary embryos, which formed within the disk lifetime, had a mass presumably smaller than that of Mars. Thus, they did not migrate significantly while in the protoplanetary disk. In addition, they did not accumulate a nebula captured primordial hydrogen atmosphere (see Lammer et al., 2014; Stökl et al., 2016). Thus, the final atmosphere of the Earth was mainly formed by degassing. However, if Earth accreted already more than 0.1 $M_{Earth}$ during the solar nebula phase (which would be in agreement with radioactive chronometers), it likely could have had a primordial hydrogen atmosphere, which was lost later on (Johnstone et al., 2015b).

Super-Earths are worlds, possibly rocky, with up to ~10 Earth masses (mini-Neptune) based on theoretical estimates for runaway gas accretion (Rafikov, 2006). Many of those found to-date orbit much closer to the central star than the Earth. Some of them may have migrated from the outer parts of the disk, thus they are probably more similar to Neptune than the Earth (rich in ice), and are sometimes referred to as mini-Neptunes. However, rocky super-Earths may have formed differently compared to our planet. In fact, when more mass is available in the inner system, namely when there is a higher flux of pebbles from the outer part of the disk, the planetary embryos grow faster and to bigger sizes. Thus, they start to be affected significantly by orbital migration. Migration in turns affects strongly the accretion process. Close-in super-Earths were already massive (e.g. Earth-mass) within the disk lifetime. Their growth was dominated by the accretion of small particles, and post-disc giant impacts may have been rare. Primordial atmospheres made of H and He are likely. The water fraction can be high for those bodies that migrated from beyond the snowline.

Earth-like planets probably have masses ≤ 5 Earth-mass (Rogers, 2015; Fulton et al., 2017). There is little direct evidence for rocky super-Earths so far in the observation of exoplanets. We have observed up to now mainly massive super-Earths or mini-Neptunes, for which initial information about interior and atmospheric properties is starting to become available (e.g., Madhusudhan et al., 2012; Wagner et al., 2012; Rauer et al., 2011). It is therefore interesting to investigate the potential habitability of super Earths from the geophysical point of view. Note that although lying in the HZ is a pre-requisite for habitability, it does not however guarantee that celestial bodies lying in this region will indeed be habitable and note that even outside the classical HZ, bodies like icy moons could be habitable as well (see discussion on the icy moons and the above definition of habitability).

The Habitable Zone (HZ) is a circumstellar region where one can find liquid surface water on an Earth-like planet under conditions of an Earth-like atmosphere. It must be noted though that for liquid water to stay long on the surface of a planet, the carbonate-silicate cycle plays a key role in regulating the climate of planets, and can prevent the occurrence of a runaway greenhouse effect (when the water content evaporates and later becomes irreversibly lost) and also can bring a planet back from a snowball state (Kasting et al., 1993a). If one takes the Earth and puts it in the HZ of an M-dwarf star, its atmosphere will change due to the interaction with the star emissions (reduced luminosity, shift in spectrum, and also occurrence of flares), which can be seen with a chemical-climate model. Coupling with a biogeochemistry model allows determining what kind of composition we can get.

Beside planets, moons can also be habitable (Kaltenegger, 2010; Heller and Barnes, 2013; Heller et al., 2014; Dobos and Turner, 2015; Forgan and Dobos, 2016; Dobos et al., 2017). In the Solar System, Enceladus and Europa (moons of Saturn and Jupiter, respectively) are the most likely hosts of extraterrestrial life but subsurface layers of Titan and possibly even Triton may have liquid water (e.g., Lunine, 2005). Below the ice shell of Europa and other ice moons, global oceans may exist, heated by tidal dissipation in the ice shell or through hydrothermal vents from the rocky deeper interior (e.g., Spohn and Schubert, 2003; Hussmann et al., 2006). In particular, in Europa's and Enceladus' oceans, a warmer environment where water interacts with the rocky seafloor may exist and simple life forms may appear. Such environments may exist in other planetary systems, as well. Moons can form similarly to planets, i.e. in the circumplanetary disk, or by collision (as it was probably the case for the Moon), or by binary-exchange which requires two bodies of comparable size that encounter a giant planet, which captures one of the binary bodies, while the other one is ejected (Barr 2016, Williams 2013).

## 2. Role of cometary, meteorite and asteroid impacts on planetary evolution

The evolution of planets and life has been influenced by collisions throughout the history of our planetary system, as evidenced by the cratered landscapes on terrestrial surfaces. The violent bombardment of the primordial planets affected their thermal evolution, and may have been crucial for the formation of habitable worlds. Volatile-rich carbonaceous chondrites may have been important sources of water and pre-biotic molecules delivering key ingredients for the formation of an atmosphere and biosphere (e.g., Albarède, 2009; Pizzarello et al., 2001), with very little cometary contribution (less than 1 % for water and carbon; Marty et al., 2016). However, the delivery of volatiles by impacts that may have significantly contributed to the growth of atmospheres is counteracted by impact-induced atmospheric erosion. The current state of research to quantify the source and loss processes due to impacts is mostly based on numerical modelling (e.g., Pham et al., 2009; Shuvalov et al., 2014). Impact erosion of the atmosphere is suggested to be more efficient for Mars compared to Venus and Earth, because of its smaller size and gravity (Pham et al., 2011). The threshold mass for ejecta escape, the threshold velocity for erosion, the threshold velocity for vaporization and the impactor obliquity factor are the major parameters to consider.

### 2.1. Impactors

Although micrometeorites (<2 mm) dominate the extra-terrestrial flux to Earth (40,000 tons/year), impacts of km-sized objects affect Earth's evolution much more strongly (Schlichting et al., 2015). Impactors with diameters from ~600 m up to 5 km – which are thought to cause global catastrophes – still occur once every 0.1 to 1 million years (Pierazzo and Artemieva, 2012). Among meteorites, the primary source of compositional information on extraterrestrial matter on Earth, chondrites represent 86% of the falls, with 80% of these related to the so-called ordinary chondrites. Chondrites are mainly composed of chondrules and Calcium-Aluminum-rich inclusions (CAIs), imbedded in a fine-grained matrix (e.g., Goderis et al., 2016). Relative to ordinary chondrites, carbonaceous chondrites (types CI, CM, CO, CV, CK, CR, CH, and CB), a subset of chondrites that makes up ~4% of meteorite falls, are enriched in refractory lithophile elements such as Ca and Al, and plot in a distinct area of the triple-oxygen isotope space (i.e., below the terrestrial fractionation line; Clayton, 2005). Despite their name, only the CM, CR, and CI groups are significantly enriched in carbon (up to a few percent as organic matter) and water relative to ordinary chondrites (Gounelle, 2011).

Most extraterrestrial matter such as water and organics on Earth is thought to be derived from asteroidal sources during the formation of the planet, with the distinction between comets and asteroids based on activity or orbital properties. Comets are active bodies that develop an extended tail (coma) when they approach the Sun close enough to sublimate water and other volatiles at their surface due to their elliptical orbits, many but not all of which extend beyond Jupiter (e.g., Gounelle, 2011; Fernandez et al., 2015). Asteroids have roughly circular orbits inside the orbit of Jupiter and do not develop comae as they do not get close enough to the Sun and may not contain sufficient volatile components at their surface. Although comets are thought to contain approximately equal amounts of dust and ice, recent sample recovery missions (e.g., Stardust) have shown that a continuum may exist between primitive, dark (with albedos of < 6%) asteroids and comets, and cometary solids and carbonaceous chondrites comparable to the difference among the different groups of carbonaceous chondrites (cf. summary in Gounelle, 2011).

## 2.2. Surface traces of impactors

Currently on Earth, approximately 190 terrestrial impact craters are known, ranging from 13.5 m to 160 km in diameter for the collapsed transient crater. This number is small compared to the Moon (which features 5185 craters with diameters ≥ 20 km; Head et al., 2010). This mainly reflects the geological activity of our planet – the number of craters found in a particular region on Earth is found to correlate with the available knowledge concerning geological activity. Since terrestrial impact structures are often affected by erosion, tectonic deformation and burial by sedimentation, their identification primarily relies on the occurrence of shock metamorphic effects or geochemical and isotopic anomalies induced by the contamination of impact melt rocks and ejecta material with meteoritic matter (e.g., Goderis et al., 2012; Koeberl et al., 2012).

These terrestrial structures provide ground truth data on the geologic effects of impacts and the subsurface structure of impact craters on other terrestrial planetary bodies (e.g., the Moon or Mars). The bombardment history of the inner Solar System is uniquely revealed on the Moon. Mass delivery due to impacts on Earth is about seventeen times higher than on the Moon due to the Earth's larger size and gravity; approximately 70 Chicxulub-sized events took place on Earth since 3.7-3.8 Ga and 15 basin-forming impactors struck Earth between 2.5 and 3.7 Ga ago (Bottke et al., 2012). The terrestrial impact history before ~2 Ga is mainly restricted to spherule layers, which are distal ejecta from large impacts (Simonson and Glass, 2004; Glass and Simonson, 2012). On Earth, the environmental effects of impact events differ with respect to time (seconds to decades) and spatial (local to global) scales. Short-term effects include ejecta blankets, thermal radiation, blast-wave propagation in the atmosphere, crater excavation, earthquakes, and tsunamis, while long-term consequences comprise of ejected dust and climate-active gases (carbon dioxide, sulfur oxides, water vapor, methane) into the atmosphere (Pierazzo and Artemieva, 2012). Impacts have thus short- and long-term consequences on the surface evolution and life. Impact cratering may not only be destructive in nature, as impact cratering may have created hydrothermal systems in the Archean (or even before) crust inducing environmental conditions ($H_2O$, heat, metals) favorable for prebiotic synthesis and perhaps organism diversification (Cockell, 2006).

In addition to the fact that impacts shaped the evolution of planets and how Earth evolved into a habitable world, the origin of life on Earth may also be a consequence of impact: the "Lithopanspermia" hypothesis considers the transfer of life-seeded rock fragments ejected from one planetary body by impact and then delivered through space to another planetary body as meteorites (see e.g. Veras et al., 2018).

Brecciation (formation by external shock of breccia, i.e. rock composed of broken fragments of minerals or rock cemented together) and impact melting of the target may have led to long-term surface and subsurface hydrothermal activity and may have contributed to possible habitats for the origin of life and its continued evolution, in particular during the early Archaean. However, large impacts also pose a significant threat for developed biospheres through catastrophic environmental consequences. For example, 66 Ma ago the Chicxulub impact event caused one of the most pronounced mass extinctions in Earth history (Schulte et al., 2010). From a chemical point of view, the extraterrestrial flux to Earth has been suggested to be an important source of bioessential elements, such as phosphorus (Pasek et al., 2013), but may also contribute considerable amounts of toxic metals (e.g., Ni; Davenport et al., 1990).

Therefore, impacts have definitely shaped the evolution of planets (see Figure 6, which will be fully explained below). It must be mentioned that impacts also induce interior changes, which is explored in Section 3.5.

# 3. Planetary interiors, mantle convection and their evolutions

## 3.1. Internal structure

After and during their accretion and subsequent melting, terrestrial planets became differentiated, leading them to possess a core and a mantle. The lithosphere can be fragmented in several mobile plates, or being mono-plate, as a stagnant lid. Heat mainly arising from accretion and present-day radioactivity is transferred from the core and through the mantle by conduction and convection. Heat transport through the colder outer layers to the surface is by conduction, as well as by convection in the case of mobile-lid plate tectonics in which the whole silicate layer up to the surface takes part in convection, and is then subsequently lost to space. This heat loss is enhanced by the existence of plate tectonics, if the latter occurs. Shortly after formation, terrestrial planets might undergo a mantle overturn because the latest cumulates solidifying at the top of the mantle are denser than the ones formed earlier at the bottom of the mantle (Elkins-Tanton et al., 2003). Isotopic compositions of some Martian meteorites can be interpreted as evidence for such an episode that occurred early in the planet's history, therewith constraining the initial evolution (Debaille et al., 2009). The occurrence of a major mantle overturn could also decrease the thermal gradient between the core and the mantle, by injecting heat-producing (rich in radioactive) elements into the deep mantle (e.g., Hess and Parmentier, 1995; Elkins-Tanton et al., 2003; Debaille et al., 2009), hence inhibiting the magnetic field.

Planetary interiors once differentiated are important for their thermal evolution. The dimension and physical states of the planet and the core are considered to be important in their evolution, possibly with or even without plate tectonics (e.g., Valencia et al., 2007; O'Neill and Lenardic, 2007; Korenaga, 2010; Van Heck and Tackley, 2011; Stamenkovic et al., 2012) and in the formation of a secondary atmosphere through magmatic outgassing that is needed to preserve surface water. Volcanic activity and associated outgassing in one-plate planets is strongly reduced after the magma ocean outgassing phase, if the planet mass and/or core-mass fraction exceeds a critical value (which depends on the mantle composition, Tosi et al., 2017; Noack et al., 2017; Dorn et al., 2017). Both degassing and volcanism are important for atmospheric evolution (see Figure 1).

The volatile elements, C-H-O-N-P-S, that are critical in defining the habitability of a planet have been processed during the internal and primordial differentiation of a planet (Southam et al., 2015; Cockell et al., 2016; Marty et al., 2016). These elements are unevenly distributed at the surface of terrestrial planets: for example, Venus lacks water (Kasting, 1988; Kasting et al., 1988; 2015), while Mars has a surface that is dominated by sulfur but strongly depleted in carbon with respect to Earth and Venus (Gaillard et al., 2012). We have in fact different compositions on different planets at various locations in the Solar System/planetary systems, which could also involve icy bodies. A very limited number of studies have addressed this issue, and most of the work is recent, implying that the state of the art is likely to significantly evolve during the next ten years. The magma ocean stage marking the early history of most terrestrial planets is believed to distribute volatile species between the planet's interior and the coexisting atmosphere (e.g., Gaillard and Scaillet, 2009; Hirschmann, 2012). The magma ocean covers a vast range of temperature and pressure and is reduced, possibly due to accretion dominated by enstatite chondrites (Javoy et al., 2010), allowing the saturation of iron from a molten mantle (Frost et al., 2008). Under such conditions, the core formation (i.e. equilibration between iron-rich metal and molten silicate) leaves behind a magma ocean, and therefore a silicate mantle, that can be depleted in carbon (Hirschmann, 2012; Li et al., 2015), hydrogen (Okuchi, 1997), nitrogen (Roskosz et al., 2013) and sulfur (Gaillard et al., 2012). The behavior of volatiles in the P-T-redox parameter space of the terrestrial magma oceans is not yet well constrained, but we know that considerable carbon has remained in both the crust and the mantle (Marty et al., 2016). This is a paradox as most of the carbon, basically a siderophile element, should have segregated into the core. Several explanations have been given in the literature, which includes carbon delivered by the impactor that has generated the Moon (Grewal et al., 2019). The effect of early planetary differentiation on the surficial volatile contents remains poorly constrained and the specific delivery process of carbon to the Earth's surface is unknown (outgassing of primordial C vs late veneer), though recent geochemical constraints limit the role of the late veneer events (Marty et al., 2016). These issues on the provenance and the fate of volatile components throughout the planetary structuration must be investigated in upcoming research programs.

After crystallization of the magma ocean, dense material cumulates just below the enriched layer and sinks to the deep interior. This mixed enriched layer encircles the core and insulates it from the rest of the mantle, trapping heat in the core (forming a "thermal blanket") and preventing the core from cooling convectively, as well as from developing a dynamo. After a certain period of time, the radiogenic material within this thermal blanket decays and heats the lower layer. With the removal of the thermal blanket, the core is then able to convect vigorously, which can produce a thermal dynamo (Breuer and Moore, 2015).

The timing and duration of the magma ocean phase and the subsequent thick, steam atmosphere (see e.g., Hayashi et al., 1979: Lammer et al., 2011) which is catastrophically outgassed during solidification of the magma ocean could potentially influence subsequent atmospheric evolution. The presence of a steam atmosphere protects the lower atmosphere from harsh incoming radiation hence influences atmospheric escape rates, photochemistry and climate during a critical early stage in Earth's history when incoming EUV fluxes were up to ~100 times higher than present day.

Figure 1 summarizes the processes affecting habitability and Heat/Interior/Atmosphere Evolution.

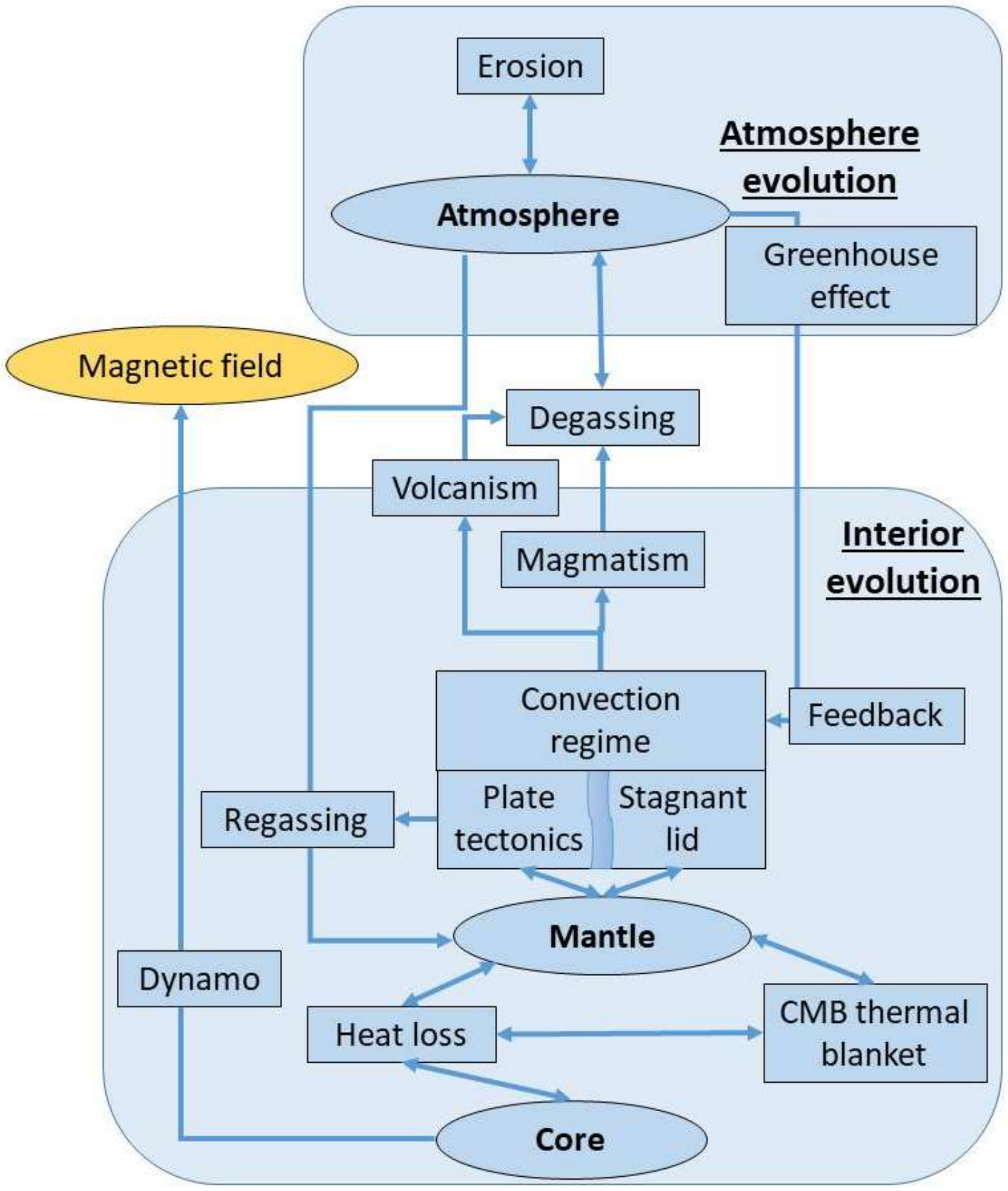

*Figure 1: Processes affecting habitability and Heat/Interior/Atmosphere Evolution.*

### 3.2. Mantle convection regime

Convection of the rocky mantle is a key process that drives the evolution of the interior: it controls heat loss from the metallic core (which ultimately generates the magnetic field) and can drive plate tectonics, which in turn governs the long-term volatile cycling between the atmosphere/ocean and the interior. Cycling of water and carbon dioxide in particular (although other species like nitrogen are also important) between the atmosphere/ocean and interior is thought to regulate habitability (e.g., Langmuir and Broecker, 2012). Indeed, the more $CO_2$ and water in the atmosphere, the higher the temperature, and the more weathering occurs, in turn regulating $CO_2$ content. Some studies suggest that planets with plate tectonics have larger outgassing rates compared with planets having stagnant lids, and that such processes could therefore be important for atmospheric formation (e.g., O'Neill et al., 2007; Noack et al., 2014). Further influence from plate tectonic on mantle convection and vice versa are provided in Section 5.3.

The comparison between Earth, Mars and Venus shows that the rocky mantle of terrestrial planets can shape their possible surface habitability via different internal processes like plate tectonics and volcanic activity. Similar feedback mechanisms between interior and surface are thought to exist on rocky exoplanets, even if they may have different chemical compositions (which might be correlated with the host star's composition) (see Elkins Tanton and Seager, 2008). Unfortunately, plate tectonics is still not well understood. It is unclear why the Earth is the only Solar System planet with plate tectonics and it is still debated when plate tectonics started on Earth. It has been proposed that Earth's plate tectonics is the remnant of magma ocean convection and thus present for most of the planet's history (Solomatov, 2016). Arndt and Nisbet (2012) present an overview of the geological and geochemical evidence for plate tectonics in the Archean. Others have argued that plate tectonics only existed on Earth since 2.5 – 3.2 Gyr (e.g., Debaille et al., 2013; Dhuime et al., 2012, 2015; Condie, 2018) with the earliest evidence so far for modern style plate tectonics at 2.2-2.1 Gyr (François et al., 2018), or possibly in place only during the last 1.8 Gyr (Gerya et al., 2015; Weller and St-Onge, 2017; Warren 2017) or 0.85 – 1 Gyr (Stern et al., 2016). In a similar way, Mars could also have experienced plate tectonics before switching to stagnant lid convection (e.g., Sleep, 1994; Nimmo and Stevenson, 2000) and Venus could have had convection regimes close to plate tectonics for some time (Gillmann and Tackley, 2014; Davaille et al., 2017).

At present-day, it is not known precisely how plate tectonics initiates and ends. It has long been held that water is the major factor in sustaining plate tectonics (e.g. Tikoo and Elkins-Tanton, 2017), because it generally lowers viscosities of mantle rock, thereby enabling subduction. Surface conditions, and in particular surface temperature, have been recognized as important factors for convection regimes (Moresi and Solomatov, 1998). High surface temperature have been reported to limit plate tectonics and favor stagnant lid due to their lowering viscosity and thus convective stress (e.g., Lenardic et al., 2008; Gillmann and Tackley, 2014). On the other hand, high enough temperatures may reduce the viscosity contrast over the upper mantle to the point where part of the lithosphere can be mobilized (Solomatov and Moresi, 1996; Noack et al., 2012). Other possible causes for plate tectonic-like behavior include melt intrusions providing weak zones in the lithosphere that can be mobilized to initiate subduction processes (Foley and Bercovici, 2014; Lourenço et al., 2017; Rozel et al., 2017) and meteoritic impacts that can mobilize a planet’s upper mantle for transient mobile episodes and provide weak zones (Gillmann et al., 2016; O’Neill et al., 2017).

### 3.3. Partial melting and the volatile pipeline

In addition to the thermomechanical processes involved in convection that are crucial to capture the magmatic production rate, there are chemical controls that must be considered in the modelling of the mantle outgassing. A key parameter is the oxidation state of the mantle, usually related to the oxygen fugacity ($fO_2$) (Frost and McCammon, 2008; Gaillard et al., 2015). In a reduced mantle, carbon remains refractory (i.e. stable as graphite or diamonds, e.g., Stagno et al., 2013; Li et al., 2017; Dasgupta and Tsuno, 2017) during melting. Reduced species such as CO or $CH_4$ (Wetzel et al., 2013; Armstrong et al., 2015; Li et al., 2017) have solubility in basaltic melts being orders of magnitudes lower than that of $CO_2$. In an oxidized mantle, in contrast, the production of $CO_2$-rich melts permits an efficient extraction of the mantle carbon (Stagno et al., 2013). We can consider that in the Earth’s mantle (which is relatively oxidized), no carbon remains in the solid residue after partial melting. Such a redox control on the efficiency of volatile outgassing from the mantle also holds for other elements such as sulfur (Jugo, 2009) and nitrogen (Li et al., 2015) in that oxidizing conditions favor efficient volatile outgassing of the mantle by melting. Which parameters control the oxidation of the Earth’s mantle, whether it varies from a geodynamic setting to another (e.g. subduction vs MORBs (mid-ocean ridge basalts); Kelley and Cottrell, 2009; Lee et al., 2010) or whether there have been major secular changes in the mantle $fO_2$ (Kasting et al., 1993b) remain unclear (Gaillard et al., 2015). Most authors (e.g., Canil, 1997; Gaillard et al., 2015; Nicklas et al., 2018) have concluded, based on geochemical proxies, that the mantle redox state has remained unchanged throughout Earth's recorded history. A recent study (Aulbach and Stagno, 2016) suggested a small secular increase in the sub-oceanic mantle $fO_2$ during the Archean, but these changes remain within the measurement uncertainties. In the absence of a definite understanding of the oxidation state of the Earth’s mantle fueling magmatism, it is difficult to evaluate how this parameter may be set in extraterrestrial settings. Yet, tentative analyses of redox systematics based on the available meteorite database suggest variations in $fO_2$ of several orders of magnitude in terrestrial bodies (Righter and Neff, 2007). An analysis of specimens from bodies with sizes ranging from small asteroids up to and including the size of Mars tends to suggest more reduced conditions than those in Earth's mantle. This systematics may be explained by disproportionation reactions in the lower mantle revealed by Frost et al. (2004) implying that planets with a lower mantle should have a more oxidized upper mantle. Wade and Wood (2005) suggested then that this might explain why the Martian upper mantle is more reduced than the Earth's one. In the upper mantle, the more consensual view is that the mantle becomes more reduced with depth: $fO_2$ tends to decrease as pressure increases due to the fact that high-pressure mantle phases are able to take on more oxygen (Frost and McCammon, 2008). This, however, must imply that the secular cooling of the Earth's mantle should produce magma that becomes more oxidized with time (Gaillard et al., 2015), while most observations indicate a steady oxidation state since the Archean. This short paragraph illustrates the contradiction between our understanding and our prediction of the redox behavior of the Earth's mantle. Oxygen fugacity of the mantle is however critical in defining the habitability of a planet. This is because it can strongly influence outgassing hence atmospheric formation, which can regulate climate, protection from radiation and the existence of liquid water on the planet's surface.

Mantle melting and volcanism define the volatile pipelines connecting the mantle to the surface where magmatism acts as a volatile cargo. In addition, the presence of volatiles in the mantle trigger melting. This feedback makes the dynamics of mantle melting more complex than usually considered (i.e. volatiles are not passively transported by melts). In particular, small amount of volatiles (i.e. the several hundred ppm of $H_2O$ and $CO_2$ usually quoted for the convective mantle) can trigger a small degree of melting (~1%) in a vast region of the mantle where seismic and magneto-telluric methods image Low Velocity Zones (Green et al., 2010; Sifré et al., 2014). These melts are very volatile-rich, but they never reach the surface for reasons which remain unclear but as such, they represent a malfunctioning of the volatile pipelines. Only unconventional geodynamic configurations can bring such volatile-rich melts to the surface as unexpectedly found in oceanic settings (e.g. Hirano et al., 2006). The role of this melting regime in the global volatile budgets and whether it is unique to Earth or can be expected on other terrestrial bodies are important issues to address.

### 3.4. Degassing, volatile cycles, and continental cycling

Mantle dynamics, volcanism and degassing processes lead to an input of gases in the atmosphere. On the other hand, volatile losses to space are estimated through atmospheric escape modeling.

Degassing under the reduced conditions which likely prevailed during the magma ocean stage (Javoy et al., 2010) can build the first atmospheres (Hirschmann, 2012) and oceans (Hamano et al., 2015; Salvador et al., 2017). Although a variety of volatile species can be dissolved in the magma ocean at high pressure (Armstrong et al., 2015; Li et al., 2017), essentially CO and $H_2$ can be produced during the low pressure equilibrium degassing from the metal-rich and reduced magma ocean (Gaillard et al., 2012). Upon cooling in the atmosphere, however, these volcanic gases must equilibrate and form $CH_4$ and $H_2O$ species.

Once the magma ocean dissipated and solidified, the $fO_2$ conditions are tentatively believed to have rapidly become oxidizing (Frost et al., 2008; Canil, 1997; Gaillard et al., 2015). Such conditions favored the outgassing of $CO_2$, $H_2O$ and nitrogen from the mantle that was precluded in the reduced magma ocean stage (Libourel et al., 2003). During the harsh conditions of the Hadean, a nitrogen-dominated atmosphere was likely not able to survive as it would have been eroded within a few million years (Lichtenegger et al., 2010; Gebauer et al., 2019) due to the high EUV flux and the strong solar winds of the early Sun (Tu et al., 2015, Johnstone et al., 2015a). A substantially higher amount of atmospheric $CO_2$ in the Hadean than at present day, however, most likely prevented strong atmospheric erosion of nitrogen; at least no footprint of a strong escape can be seen in the atmospheric $^{14}N/^{15}N$ isotope structure (Cartigny and Marty, 2013). Partial pressure of $CO_2$ declined during the Hadean and Archean (e.g., Hessler et al., 2004; Kanzaki and Murakami, 2015). This suggests a $CO_2$-dominated atmosphere during the late Hadean $N_2$, which, at some point, likely became the main constituent due to efficient outgassing from the oxidized mantle (Mikhail and Sverjensky, 2014; Lammer et al., 2018). The mid-Archean Nitrogen and Argon isotope data from atmospheric inclusions in hydrothermal quartz fluids indicate that the partial pressure of $N_2$ was 0.5-1.1 bar for 3.0-3.5 Gyr ago (Marty et al., 2013) or similar or lower than the modern one for a similar time span (Avice et al., 2018), whereas about 2.7 Gyr ago the total surface pressure was found to only be 0.23 ±0.23 bar (Som et al., 2016). This potential decline in atmospheric nitrogen is likely to be due to the onset of biogenic nitrogen fixation at least 3.2 Gyr ago (Stüeken et al., 2015; Stüeken et al., 2016b; Zerkle and Mikhail, 2017). The onset of biotic nitrogen-releasing processes such as denitrification during the Great Oxidation Event (GOE) ~2.4 Gyr ago on the other hand served as an additional nitrogen source counterbalancing abiotic and biotic fixation processes (Zerkle and Mikhail, 2017; Lammer et al., 2019). The GOE thus marks the beginning of the modern geobiological nitrogen cycle (Zerkle et al., 2017).

As mentioned in Section 3.3, $CO_2$ degassing is much more efficient on Earth than on Mars due to likely much more oxidizing conditions in the post-magma-ocean Earth's mantle than its Martian counterpart (e.g., Stagno et al., 2013 vs. Grott et al., 2011). The contrasting mantle $fO_2$ conditions between both planetary bodies is certainly linked to planetary size (Wade and Wood, 2005). Nevertheless, the $H_2O/CO_2$ ratio of volcanic gases during that era was controlled by the atmospheric pressure, with dense atmospheres favoring dry volcanic emissions while tenuous atmospheres favored $H_2O$-rich volcanic gases (Gaillard and Scaillet, 2014; Tosi et al., 2017). This is linked to the fact that $H_2O$ is orders of magnitude more soluble than $CO_2$ in basalts and that solubility laws imply that, with decreasing pressure, the faction of water partitioning into the fluid must increase (see Iacono-Marziano et al., 2012). This results in a dilution of $CO_2$ in the magmatic fluids. Many geochemical observations on volcanic centers attest this behavior (e.g. Edmonds and Gerlach, 2007). We nevertheless must acknowledge that this is true in a reasonable range of $CO_2$ and $H_2O$ contents in the magmatic source. If a mantle contains large amount of water and very little $CO_2$, this degassing filter will be poorly efficient and the volcanic gas compositions will mostly reflect the composition of the magmatic mantle sources. Noteworthy, though this has never been specifically reviewed in the literature, this filtering process by volcanic degassing also applies to sulfur degassing (sulfur-rich magma may not degas sulfur-rich gas). All this complicates the planetary degassing picture and introduces feedbacks that seem promising to address and to relate to the concept of formation of the oceans, climate and habitability (Gaillard and Scaillet, 2014; Tosi et al., 2017).

Once the oceans are formed, submarine volcanism and subaerial volcanic degassing must be taken into account and both processes may lead to very different compositions of volcanic gases (Gaillard et al., 2011). For example, sulfur (and water) outgassing is mostly not possible at seafloor pressure implying that submarine volcanism does not supply $SO_2$ to the biosphere while subaerial volcanic gases are very $SO_2$-rich. Historically, some confusion has arisen in the planetary sciences literature due to the assumption that seafloor volcanism and hydrothermal venting are identical processes (Kump and Barley, 2007). While the former is recognized to outgas mantle sulfur less efficiently (Gaillard et al., 2011), the latter, well known as black smokers, emits massive quantity of $H_2S$ and FeS in the ocean. The sulfur emitted by black smokers is however not of mantle origin since it mostly recycles seawater sulfates by sulfate reduction processes (Holland, 2002; Lyons and Gill, 2009). It appears therefore that only subaerial volcanism can produce $SO_2$-rich volcanic gases. A fraction of outgassed $SO_2$ can be oxidized to $SO_3$ depending on the redox state of the atmosphere. $SO_2$ plays a role in generating mass-independently fractionated sulfur isotopes on the Early Earth (e.g., Farquhar et al., 2000; Pavlov and Kasting, 2002). The whole chain of processes ruling the sulfur biogeochemical cycles, as we know it on present-day Earth, may therefore become rate-limited by volcanic outgassing of $SO_2$ if submarine volcanism dominates on a planetary body. Given the importance of the sulfate-reduction reactions associated with atmospheric $SO_2$ (Halevy et al., 2010), the ratio of submarine to subaerial volcanism is an important parameter to consider.

At the same time, the volatile content of the surface environment, particularly the presence of liquid water is thought to have a large feedback on the interior, for example by favoring plate tectonics (e.g. Nakagawa and Iwamori, 2017). Partial melting depletes the mantle in volatiles and extracts water from the interior to the surface, where it is released via volcanism (e.g., Breuer et al., 2016 for a review for Mars; and Poirier, 2000; Karato et al., 2013, for Earth's interior). Volatiles can accumulate in a planet's atmosphere but other mechanisms exist that can remove them from it. Those so-called loss mechanisms can be very diverse but can be classified into three different types and are detailed in other sections of this manuscript:

i) Atmospheric escape that covers interaction between radiation by the sun and the upper atmosphere, leading to the loss of volatile into space (Tian, 2015). It is generally most efficient early in the evolution of the planet. Large Impacts or cumulative effect of small impacts could lead to atmospheric erosion (See Section 4.2).
ii) Interaction with the interior of the solid planet, like silicate rock alteration, carbonate formation and seafloor sequestration before possible subduction (Walker et al., 1981; Kump et al., 2000). Part of the volatile cycle, it operates on a million years timescale.
iii) Other singular processes that can be transient (like the formation of a polar cap, which does not affect the global volatile budget) or remove gases for good (like impact erosion).

Over time, those escape mechanisms can substantially modify atmospheric mass but also composition, in cases where species are removed preferentially (like the light species from atmospheric escape). Such processes can be studied by analyzing the isotopic fractionation that they cause and are believed to have potentially far reaching consequences. They could even govern the atmospheric evolution if enough water remain in the atmosphere to condensate into an ocean for example (Gillmann et al., 2009; Hamano et al., 2015). Outgassing and volcanism are also related to volatile cycles. It is thus necessary to consider a coupled atmosphere-interior evolution for the understanding of habitability (see Figure 1).

The existence of plate tectonics leads to considerably more complex carbon and water cycles, which could significantly alter the outgassing expected from a convective mantle covered by a lid: an unknown amount of carbon, hydrogen, and other atmophile elements is reintroduced into the mantle with the oceanic lithosphere at subduction zone. The sediments can carry a considerable amount of carbon and water (Plank, 2014) and fluid-rock reactions occurring at seafloor hydrate and introduce carbonates into the oceanic crust down to a depth that remains debated (Alt et al., 2012). This form of C-H atmosphere-lithosphere exchange is related to the presence of the water cycle at the surface. Subduction geometries and thermal status imply then that the surficial carbon and hydrogen must be returned to the convective mantle and sustain a long-term geodynamic cycle (Hammouda and Keshav, 2015; Höning and Spohn, 2016). Capturing fluxes of carbon and hydrogen associated with subduction and how these compare with the outgassing rates of C and H from volcanic centers is a central task which is difficult to quantify (Hirschmann and Dasgupta, 2009; Kelemen and Manning, 2015) but which is strongly relevant to the understanding of habitable world.

When water is incorporated into the Earth's mantle, it is believed to reduce its viscosity (e.g., Karato et al., 1986). Since a reduction in viscosity enhances both the water degassing and regassing rates, the mantle water budget can be described as a feedback cycle (e.g., McGovern and Schubert, 1989). In addition, subduction of water is crucial to the production of continents (e.g., Campbell and Taylor, 1983). With the emergence of a large continental area, the weathering rates increase. As a result, large amounts of sediments are produced and subducted into the Earth's mantle. Subducting sediments carrying water in stable phases complete the continental feedback cycle thereby further coupling it to the mantle water cycle (illustrated in Figure 2).

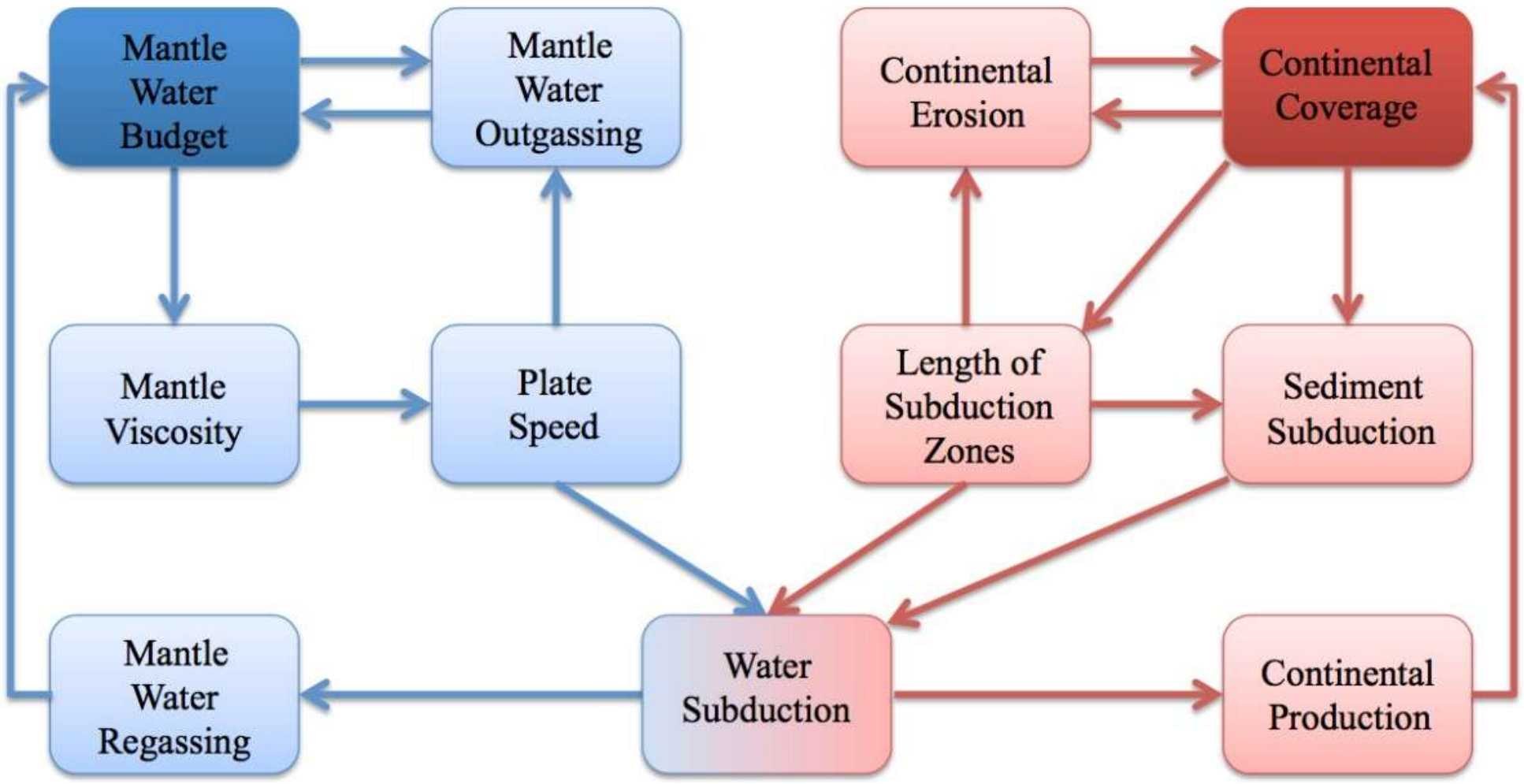

*Figure 2: Coupled feedback cycles related to mantle water budget (blue) and continental coverage (red), adapted from Höning and Spohn (2016).*

Particularly important are self-reinforcing mechanisms associated with continental growth, such as an increased subduction rate of sediments with the emergence of continents. Although these feedbacks are difficult to quantify, a system with strong positive feedbacks is well suited to explain the geological record of continental growth (Höning and Spohn, 2016). This implies that multiple equilibrium points could exist in the continental and mantle water system (Höning and Spohn, 2016), implying that the present-day Earth fraction of emerged continents is not a necessary result for Earth-sized plate tectonic planets in general. Rather, the fraction of emerged continents should strongly depend on initial conditions (e.g., initial mantle temperature, initiation time of plate tectonics) as well as on the weathering rate (Höning et al., 2016; 2019).

## 3.5. Consequences of impacts on the interior of planets

Impacts of cosmic bodies on planetary surfaces cause consequences in several respects. Most obvious is the formation of a crater, but other effects such as fracturing of crustal rocks, ejection of material, heating of crustal and mantle rocks up to melting and vaporization temperatures, and the addition of material to the target body are of crucial importance.

Consequences of impacts on solid planetary surfaces have been investigated by laboratory analogue experiments (e.g. Kenkmann et al., 2017) and numerical modeling (e.g. Collins et al., 2013; Jutzi et al., 2015). Impacts forming craters smaller than some tenth of kilometers in diameter are considered to have little direct effect on the long-term evolution of a planet; however, such impacts considerably contribute to the development of planetary landscapes that are often peppered by thousands of craters varying in size and frequency, with small craters occurring much more frequent than larger crater structures. The size-frequency distribution of craters provides a way to date the planet's surface (e.g. Werner and Ivanov, 2015) and to estimate its erosive, volcanic (resurfacing) and interior activity.

Upon hypervelocity impact a shock wave is generated that propagates hemispherically away from the point of impact decreasing in amplitude as a function of distance (e.g., O'Keefe and Ahrens, 1977; Melosh, 1989). As a consequence of shock wave compression and subsequent unloading material is heated (the temperature increase is directly proportional to the maximum shock pressure $P_{max}$ the material experienced, e.g. Pierazzo et al., 1997) and accelerated. The former causes vaporization ($P_{max}$>100-150 GPa) melting ($P_{max}$>~60 GPa), or an increase in temperature, the latter is the driving process resulting in the excavation of a crater by displacement and ejection of material. The ejecta plume is composed of vapor, molten, and lithic particles originating from the impactor, but mostly from the target. Lithic ejecta vary significantly in size ranging from dust particles to large fragments whose upper size limit depends on the size of the crater. The interaction of the ejecta plume with an existing atmosphere has constructive and destructive consequences for the evolution of atmospheres, which is discussed in the next section (Section 4.2). A small amount of material is ejected sufficiently fast to escape the planet's gravity field and to fall as meteorite on other planets such as Martian meteorites on Earth (Head et al., 2002; Artemieva and Ivanov, 2004). By this mechanism, an interplanetary exchange of mass and maybe also life (lithopanspermia) is possible (Napier, 2004; Cockell and Cockell, 2008; Cockell et al., 2015; Veras et al., 2018). The vast majority of the ejected material follows approximately ballistic trajectories and is deposited as a continuous ejecta layer up to 2-3 crater radii away from the crater. The thickness of the layer decreases approximately proportional to a power-law with distance and can reach a thickness of several kilometers close to the crater for impact basins several 100 km in diameter. Impactors forming basin structures of such a size (impactor radius>100km) lead to the redistribution of mass several times the impactor mass over the whole planet (Shuvalov et al., 2012). The accumulative effect of ejecta deposition by numerous impact events ranging from impact craters of a few meters to several 10s of kilometers and the periodically occurring complete overturn of the deposit by large basin-forming impacts are commonly referred to as "impact gardening". The resulting thick surface veneer is composed of megaregolith and regolith that grades from one into another. As such, lithological units are thought to have very poor thermal conductivity properties they have been suggested to have a strong insulation effect on the thermal history of a planet (Rolf et al., 2016).

The vast majority of the kinetic energy of the impactor is transferred into internal energy that is distributed by the shock wave deep into the interior of a planet. Depending on the size ratio of the impactor diameter and the planets diameter an increase of temperature can also occur at the antipode to the point of impact (Bierhaus et al., 2012). However, most modelling attempts to quantify the effect of impact-induced heating on the thermal evolution (e.g., Watters et al., 2009; Roberts and Arkani-Hamed, 2014; Gillmann et al., 2016., Monteux et al., 2007) often rely on simple scaling laws that are based on small-scale laboratory experiments (O'Keefe and Ahrens, 1979; Housen et al., 1983) or simple estimates from numerical modelling (e.g., Abramov et al., 2012; Pierazzo et al., 1997). Such scaling laws suggest that impacts create a thermal anomaly with an almost constant temperature increase in the near field, the so-called "isobaric core", which extends, depending on incident angle and velocity, to approximately 1-2 times the size of the impactor. Outside the isobaric core, the shock wave-induced temperature rise decreases with distance approximately proportional to the power of 2. (Pierazzo et al., 1997; Monteux et al., 2007, Ruedas, 2017). Such simple estimates are not applicable for giant collisions of planetary embryos, such as the Moon-forming event (see Nakajima and Stevenson, 2014; 2015; Cameron and Ward, 1976; Canup, 2004; Melosh, 1990; Pahlevan and Stevenson, 2007: for further details); they become already inaccurate for sufficiently large impactors (e.g. some 10s of km diameter for the Earth), where the pre-impact temperature of the interior and the interaction of the shock wave with the core-mantle boundary become important for the distribution of impact-generated heat (Manske et al., 2018).

Depending on the lithology, whether more crustal rocks or mantle material is involved, a melt volume is generated that corresponds approximately 5-15 times the volume of the projectile; however, this does include the additional melting that occurs due to decompression melting as a consequence of the uplift of originally deep–seated mantle rocks (Manske et al., 2018; Jones et al., 2005).

Besides the immediate production of shock melt, the impact-induced heating of the interior of a planet has further important implications. Smaller impacts generate shallow thermal anomalies that do not reach far enough to affect the mantle, but cause melting and degassing of the lithosphere (see Section 4.2). Impactors big enough so that their thermal anomaly significantly reaches into the mantle, can have long-term and sustainable consequences for convection processes. The impact-induced thermal anomaly is hot and positively buoyant. It rises toward and flattens against the surface, leading to lateral spreading in a low viscosity flow. The material can stay hot for hundreds of thousands to millions of years, leading to widespread melting and resurfacing (Watters et al., 2009; Roberts and Arkani-Hamed, 2014; Gillmann et al., 2016; O'Neill et al., 2017; Rolf et al., 2016, Padovan et al., 2017; Ruedas and Breuer, 2017). For major impacts, the thermal anomaly can spread beyond the hemisphere the impact occurred on, leading to global scale perturbations. To conserve mass, the upper mantle is mobilized, with material being pushed away from the impact location. Eventually, it may accumulate at an antipodal position, where it forms downwellings/subduction events (Gillmann et al., 2016; O'Neill et al., 2017). In this situation, large impacts can drive a transient plate tectonics-like global convection in the mantle of terrestrial planets for tens of millions of years. Moderately sized impacts (10 km < radius < 100 km) can help triggering these events by thinning lithosphere causing upwelling.

Mantle partial melting also means mantle depletion. Water and $CO_2$, especially, are incompatible species and go into the melt easily. Thus, impacts could contribute to remove volatiles from planetary mantles (Davies, 2008). It has been suggested that a single large impact (radius>800km) may be capable to efficiently deplete the upper mantle in the vicinity of the impact due to the large extent of the thermal anomaly generated by such an event (Gillmann et al., 2016). However a number of moderately large impacts (200km<radius<500km) is much more efficient to deplete the upper mantle of a planet than a single event of the same cumulated mass (Gillmann et al., 2017). Such a scenario leads to a drier planet, especially if no rehydration (and remixing) mechanism is present (i.e. no subduction). This diminishes the odds that the planet could sustain significant degassing (especially of water) during later phases of its evolution, leading to drier, cooler surface conditions.

On the long term, large impacts can still affect the convection regime of terrestrial planets hundreds of millions of years after the event occurred. They contributed to emplace a large amount of fresh crust at the surface, which is preferentially located near the impact location, where the bulk of the melting takes place (Nimmo et al., 2008, Golabek et al., 2011, Padovan et al., 2017). Impact angle, however, has been suggested to be a crucial parameter determining crustal distribution (Golabek et al., 2018).

Sufficiently large impacts can also affect the core of a planet. For larger collisions, the thermal anomaly can reach down to the core-mantle boundary (CMB) affecting the heat flux from the core into the mantle. Large variations of heat flux (at the top but also at the bottom of the mantle) and corresponding changes in convective regime (considering massive hotspot near the top of the core or equally large-scale downwellings of colder material to the CMB) are likely to elicit a response. It is still difficult to assess the precise consequences of a large impact on the core. Roberts and Arkani-Hamed (2014) predict that the high temperature layer on top of the core would contribute to its symmetric stratification, preventing cooling of the core. As a consequence, the reduced heat flow would shut down core convection and cripple the magnetic dynamo. It could take around 1 Gyr for the core to go back to its previous convective state. On the other hand, however, O'Neill et al. (2017) suggest that an increase in the convective vigor just after an impact could lead to a resurgence of the magnetic field.

The last main effect of collisions is the delivery of material to the solid planet. During the end of the accretion, the so-called Late Veneer phase, while the bulk of the mass of terrestrial planets is already in place, it has been estimated that 0.5% of Earth's mass is still being delivered by a small number of large (radius>750 km) bodies (Raymond et al., 2013). Those impacts are thought to be responsible for the repartition of the highly siderophile elements (HSEs: Os, Ir, Ru, Rh, Pt, Pd, Re, Au; Walker, 2009), which are found in higher concentration in the mantle than would be the case in a silicate-metal equilibrium. The common explanation is that accretion went on after core formation (Kimura et al., 1974; Rubie et al., 2015). Impacts responsible for this late accretion would have been also susceptible to have a strong effect on volatile repartition and mantle convection. It is however difficult to estimate how much of the projectile mixes with the target body, especially for differentiated impactors, as would probably be the case for the late Veneer collisions. While undifferentiated material could equilibrate with the mantle, for differentiated bodies, the metallic core could sink toward the CMB with little equilibration (Kendall and Melosh, 2016). This implies that more impact mass might be needed to reach measured HSE mantle concentrations (Marchi et al., 2018).

## 4. Processes affecting atmospheric development

Atmospheric development is affected mainly by five processes, namely escape (4.1), delivery (or/and erosion) via impacts (4.2), outgassing and surface emissions (4.3), surface removal (4.4), and finally, life itself. Figure 2 summarizes these processes and their main interactions (note that the effects of life are discussed separately further below (5.0).

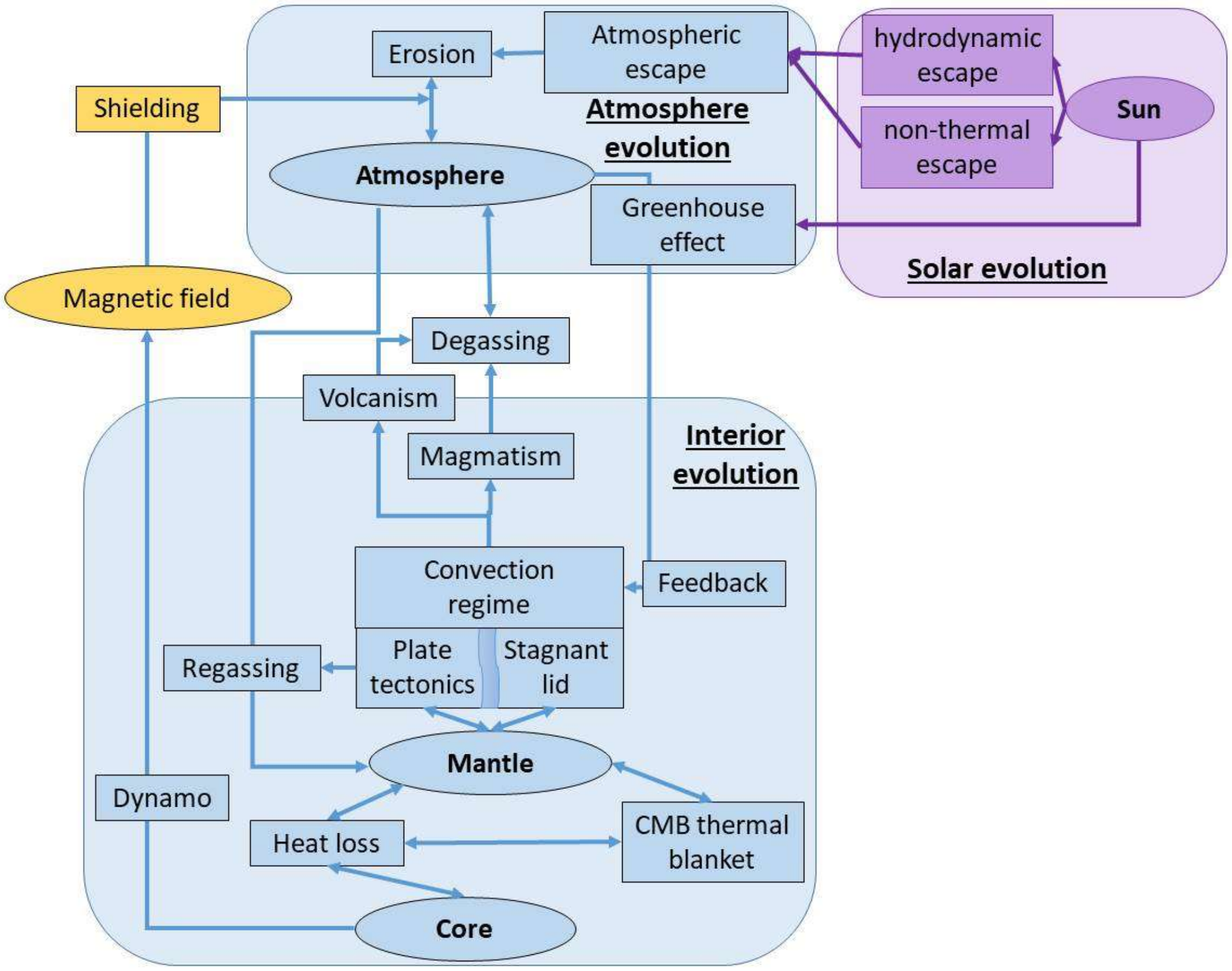

*Figure 3: Habitability and heat/interior/atmosphere evolution, and the role of the Sun and atmospheric escape*

## 4.1. Atmospheric escape

Atmospheric escape can be diffusion-limited or energy-limited depending on numerous factors such as atmospheric mass, composition, and the incoming stellar energy (see e.g. Watson et al., 1981). Also potentially relevant is the efficiency by which photochemistry can release the light hydrogen atoms from their heavier molecular reservoirs in the middle-to-upper atmosphere. Energy-limited escape includes thermal (Jeans) escape that is related to the Maxwell-Boltzmann energy distribution, so depends upon temperature hence upon atmospheric absorption and incoming stellar energy. At higher (EUV) energies, e.g. typically during the earlier stages of evolution for Earth-like planets, statistically driven Jeans escape can transition into an organized, bulk gas outflow known as hydrodynamic escape or blow-off. Larger species (such as helium, carbon or oxygen atoms) can undergo escape via so-called hydrodynamic drag where they are carried along in the stream of lighter, hydrogen atoms. Hydrodynamic escape is a determining factor in early planetary atmospheric evolution (Tian and Ida, 2015). It is believed that hydrodynamic escape is able to generate noble gas isotopic fractionation, as observed on Venus (Gillmann et al., 2009). It has however been shown (Odert et al., 2018) that non-unique set of conditions existed to reach the present-day observation. Also for close-in extrasolar planets, hydrodynamic escape is believed to play a major role in shaping the bi-modal distribution of terrestrial-type and gas-rich planets (Fulton et al., 2017; Owen and Wu, 2017). Hydrodynamic escape is an important process for establishing volatile inventories, particularly for small bodies (<0.8$M_E$, where $M_E$ is Earth's mass) as they would be unable to retain their inventory of volatiles like water (Lammer et al., 2014; 2018). Therefore, hydrodynamic or thermal escape also leads to the loss of volatiles from planetary embryos (Tian et al., 2018), thus affecting planetary formation.

Non-thermal (non-Jeans) escape processes include interactions between the atmospheric molecules and solar radiation (EUV) like charge exchange and dissociative recombination (see e.g. Shizgal and Arkos, 1996), sputtering (Chassefière and Leblanc, 2004) or plasma clouds instabilities (Lammer et al., 2006). In the present-day Solar System, non-thermal escape mechanisms constitute the bulk of the measured escape on terrestrial planets.

Non-thermal escape is thought to have had a significant effect on the long-term evolution of terrestrial planets, contributing to the shift from a wet early Mars to dry present-day conditions as evidenced by the D/H ratio in the Martian atmosphere (Carr, 1990). The same scenario has been discussed for Venus (Donahue, 1999) but here the situation is more complex (Grinspoon, 1993; 2013) because of large variations of D/H ratio with altitude (Marcq et al., 2018).

Non-thermal escape, unlike Jeans escape, is affected by planetary magnetic fields generated by core dynamos (e.g., Lammer et al., 2009). Intrinsic magnetic fields are generated by conductive fluid motion within a terrestrial planet's liquid outer core through either thermal or chemical convection. Core conditions necessary to magnetic field generation are likely highly dependent on core composition, global mantle dynamics (e.g., existence of plate tectonics), water repartition and even surface conditions (Lammer et al., 2009). Venus shows no sign of past or present-day intrinsic magnetic field due to its young surface. However, it is believed that Venus is likely to have generated one in the past even though reasons behind its absence at present-day are still debated (Stevenson et al., 1983; Jacobson et al., 2017; Dumoulin et al., 2017; O'Rourke et al., 2018). Mars, while also lacking a present-day intrinsic field, still exhibits remnants of magnetized crust (Acuña et al., 2001; Connerney et al., 2001). Differences between Mars, Venus and Earth have been linked to the lack or presence of magnetic fields (see Lammer et al., 2009), implying that water loss and cessation of habitability were related to the lack of a shielding magnetic field (Lundin et al., 2007).

Present-day escape on Venus and Mars has been measured by Venus Express, Mars Express and MAVEN. Observations suggest that escape rates for both planets is similar to Earth's despite the Earth's magnetosphere possibly acting as a shield. It has been proposed that a large magnetosphere presented a larger interaction region to the solar radiation (Barabash et al., 2007), resulting in a similar net loss. The debate is still currently open. Uncertainties about the re-entry of ions along the magnetic lines (Tarduno et al., 2014) may indicate lower actual losses and favour the shielding effect. On the other hand, some recent models indicate that losses from magnetospheres could be enhanced near the magnetic poles and cusps (Gunell et al., 2018) thus limiting further protection offered by magnetic fields.

One needs also to consider variations in conditions (both Solar and atmospheric) over the history of planets. Magnetic fields are likely to play a stronger role early on, when solar activity was stronger (Lammer et al., 2009; 2018). This affects two important aspects. First, the response of early atmospheres to Coronal Mass Ejections (CMEs) from the young Sun, much more frequent than at present-day (Lillis et al., 2015; Ramstad et al., 2017), could lead to enhanced escape on planets lacking a magnetic field. Second, a magnetic field could have played a role against large non-thermal escape rates of nitrogen caused by the high EUV flux of young stars (Lammer et al., 2018), leading to one of the defining features of Earth.

Finally, whether or not magnetic fields affect atmospheric escape, they still offer a protection to life from high-energy particles, which is perhaps the most their straightforward effect on habitability.

## 4.2. Impacts

Collisions are complex events that can affect a terrestrial planet's atmosphere in dramatic ways immediately after an impact but also have lasting long-term consequences.

The first immediate effect of impacts on the atmosphere is the heat that is released in the gaseous envelope of a planet. Even smaller impacts (Radius<50km) can heat significantly a planet's atmosphere. In some cases, the sequence of processes related to the atmospheric interaction and long-term climatic consequences may have led to extinction events as it happened 65 Ma ago on Earth. In contrast, smaller impact events (Radius<5km) may have created favorable spots for life (for example warm ponds and hydrothermal system; Sleep et al., 2001). Those impacts can also change the chemistry of the atmosphere. Bulk species ($H_2O$, $CO_2$, $SO_2$) are obviously important, but less common species are also important to life and impacts can affect their chemistry (Kasting, 1990). Iron meteorite could lead to increased concentrations of reduced gases like $H_2$ and $CH_4$, which is important for prebiotic chemistry. Organic molecules/hydrocarbons can also be generated during an impact (Muhkin et al., 1989). It is unlikely that those impacts can affect the long-term climate of a terrestrial planet. Studies have shown on Mars (Turbet et al., 2017) that single impacts (Radius<50km) do not produce stable warm climate but only local precipitations for no more than 5-7 years per bar of water injected.

Larger collisions could have more violent consequences and generate enough heat to vaporize an ocean or even create a local (or global for giant impacts) magma ocean. In an ocean-vaporizing impact, it is likely that the atmosphere goes through a transient silicate vapor phase lasting from a few hours to months, depending on the energy of the collision. This is caused by the plume rising after impact and condensing into nanoscale particles during its expansion and cooling by radiating out to space and into the ocean (Nisbet et al., 2007), leading to the formation of a steam atmosphere. The temperature of the atmosphere also decreases at the end of the phase over timescales of the order of a year, while steam atmosphere rains back down over decades to centuries after the event (Zahnle and Sleep, 1997; 2006).

The main long-term consequence of collisions is the changes they can cause to the global volatile inventory of the atmosphere. Impacts by meteorites have long been known to be able to erode planetary atmospheres (Cameron, 1983; Melosh and Vickery, 1989). Planets with lower mass (like Mars compared to Earth or Venus) are more likely to see their atmospheres eroded efficiently (Pham et al., 2011). Several different mechanisms can lead to atmospheric loss: (1) direct burst and ejection (quite substantial) of the atmosphere (related to the shock wave); (2) plume effect involving the vaporized projectile and sediments; (3) basement clasts particle ejections; the particles ejected in the atmosphere accelerate and heat the atmosphere; here the impact angle is important for the amount of particles; (4) a fourth mechanism also exists in the case of large collisions: the ground motion of the planet caused by the impact that can accelerate particles of the atmosphere above escape velocities.

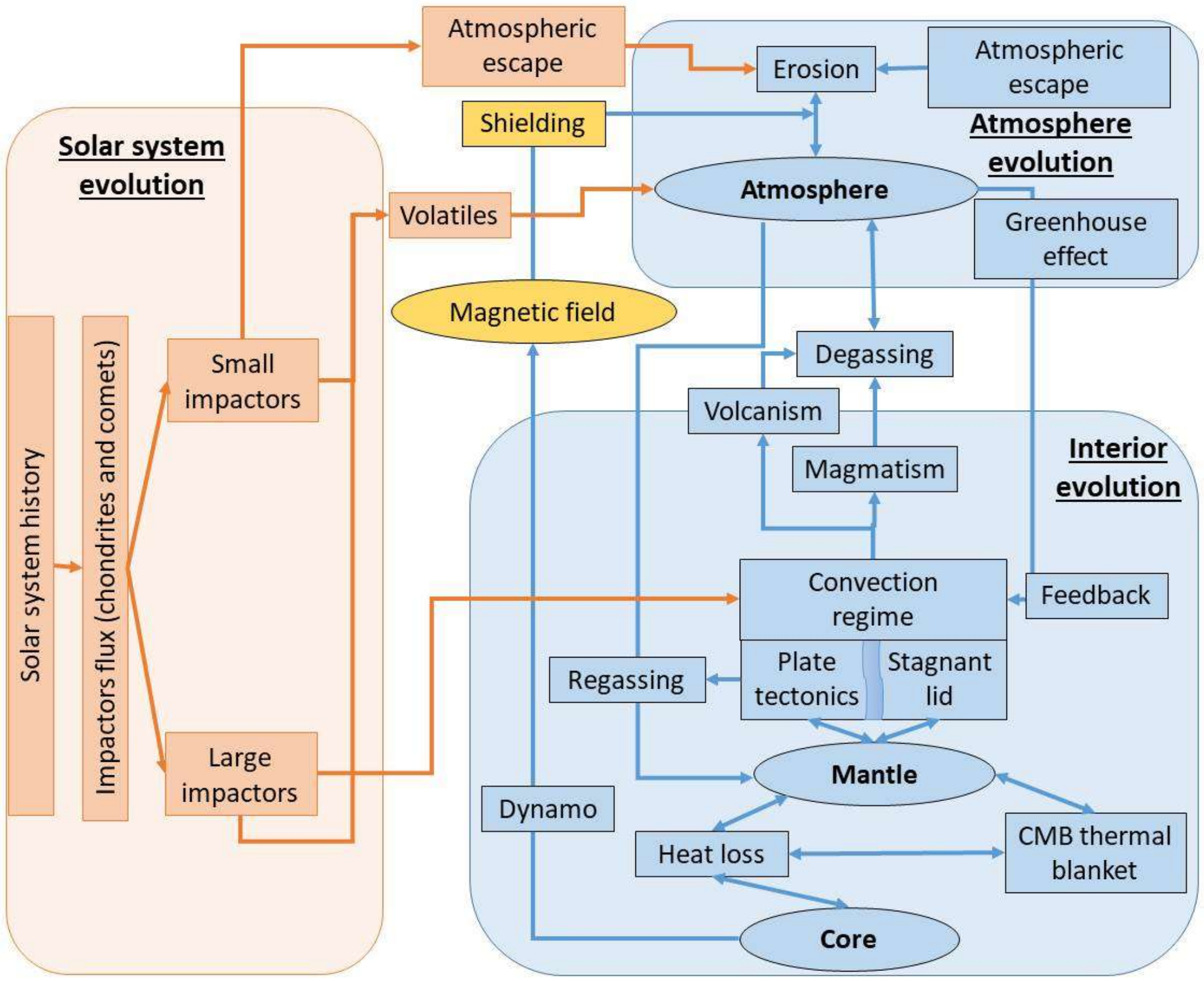

*Figure 4: Impact effects on the planet and its atmosphere.*

Atmosphere losses due to giant impacts (above 1000 km radius) are at the level of 20% of the mass of the impactor (Genda and Abe, 2003), which can be significant. Ground motion seems to be the cause of this large atmosphere loss in the giant impact scenario. Precise modeling is, however, difficult to constrain for those sizes and numerical simulations usually focus on the consequences for the solid planet. Erosion of the atmosphere by a few large impacts (above 200 km radius) appears limited, unless airburst processes are considered, especially in a dense atmosphere (Shuvalov, 2009; Shuvalov et al., 2014). Swarms of smaller (1km<R<100km) more mass-effective impactors seem required for this effect to be significant. In fact, those small impacts erode atmospheres very efficiently because most of their energy and momentum can directly be available to fuel volatile escape at a local scale. On the contrary, large/giant impact must act on a global scale and generate enough shock to generate ground motion (strong vibrations of the planet's surface), which is much less efficient. The cumulated effect of the different impactor populations is then vastly tilted in favor of small objects, mainly due to how many of those smaller bodies can affect the planet over its history (Schlichting et al., 2015; Schlichting and Mukhopadhyay, 2018).

Impactors can create a large input of volatile in the atmosphere from the melting they cause during the impact and through the volatiles that they carry. It is quite difficult to estimate which of those two origins for volatile input dominates, as it depends on many parameters, like the vaporization of the projectile, its content, target material composition and so on. However, for volatile rich meteorites (like carbonaceous chondrites for example) and comets, input from the impactor seems the major component. It is thus possible to compute the net balance of erosion and retention assuming a given impactor flux/different scenarios (i.e. Pham and Karatekin, 2016, Schlichting et al., 2015) The way impactors can deliver volatiles mainly depends on three parameters, which are the volatile content of the impactor, the amount of impactor material remaining on the planet, and the fraction of the retained impactor mass which is released as vapor. For typical asteroidal impacts, silicates are not completely vaporized (Pierazzo and Melosh, 2000). Studies on the atmospheric loss and delivery due to impacts differ sometimes by orders of magnitude, mainly due to different equation of state and dynamical models used (Hamano and Abe, 2010; Melosh and Vickery, 1989; Newman et al., 1999; Shuvalov, 2009; Svetsov, 2007; Vickery and Melosh, 1990).

This leads to an increase/decrease in atmosphere density, greenhouse effect and surface temperatures, depending on which effect is dominant. Results typically vary from planet to planet: planetary mass, atmosphere density, impactor parameters all interact to the end-result difficult to intuit (Pham et al., 2011; Marounina et al., 2015; Schlichting and Mukhopadhyay, 2018).

The vaporization rate depends on the impact energy. Smaller and slower collisions tend to contribute less to the atmosphere. It is quite difficult to estimate which of the two origins (target melting or impactor vaporization) for volatile input dominates, considering how many parameters can vary and their uncertainties. Small impactors are able to deliver volatiles on vaporization but are usually less efficient at creating degassing of the planet. For volatile-rich meteorites (like carbonaceous chondrites for example) and comets, input from the impactor seems the major component.

### 4.3. Outgassing and Surface Emissions

Outgassing of volatile species such as $CO_2$, $H_2O$, CO and Sulphur-containing compounds (see e.g., review by Cadle, 1980; Holland, 1984; Catling and Kasting, 2017) likely played an important role in shaping Earth's atmosphere. The main points have been discussed in Section 3. In addition to volcanic outgassing, surface emissions on Earth can also arise due to e.g. geological processes such as serpentinization, anthropogenic emissions e.g. from industry, vehicles etc. and biomass emissions due to life (Section 5).

### 4.4. Surface removal

Gases can be removed onto surfaces essentially depending on species properties (e.g. gas velocity, polarity etc.), the type of surface (e.g. phase, composition, roughness etc.) and atmospheric properties (e.g. density, temperature). Dry deposition is usually parameterized with a so-called (species-dependent) sticking coefficient multiplied by the flux of gas-phase species impinging onto the surface. Removal from the gas-phase into bulk surface liquids (e.g. absorption of $CO_2$ by the oceans) can be parameterized using Henry's law coefficients. Raindrops falling through the atmosphere can scavenge soluble gases and lead to gas-phase removal by rainout. For a review of these processes, see e.g., Seinfeld and Pandis (2016).

## 5. Interaction of life with the atmosphere, geosphere and interior of planets

### 5.1. Evolution of Earth's atmosphere

Earth's ($N_2$-$O_2$) dominated atmosphere is essential for all life on Earth because it regulates climate (e.g. due to radiative absorption, emission and scattering) and by distributing energy, moisture and momentum via dynamical transport. The atmosphere also offers protection from both harmful electromagnetic radiation (e.g. via UV-absorption in the ozone layer) as well as high energy particles, and most importantly, its influence upon pressure and temperature enables the long-term presence of surface liquid water, required by all life on our planet. As explained in Section 3.4, simulations however of the terrestrial paleo-magnetosphere as well as of the solar wind induced atmospheric ion-pickup escape ~4 billion years ago (Scherf et al., 2019, in prep., Lichtenegger et al., 2010) are indicating that during the harsh conditions of the Hadean and early Archean eons, a nitrogen-dominated atmosphere would not have been able to survive, but would have been eroded within a few million years due to the high EUV flux and the strong solar wind of the early Sun (as discussed above). In addition, these results are indicating higher amounts of $CO_2$ during the Hadean and early Archean in agreement with studies of the $CO_2$ partial pressure evolution (e.g., Hessler et al., 2004; Kanzaki and Murakami, 2015). Recent works have suggested that uncertainties in the early rotation rate and associated activity evolution of the Early Sun (Johnstone et al., 2015b) could have a potentially large influence on Earth's atmospheric evolution.

Early Earth (before ~2.3 Gyr) atmosphere was likely anoxic and methane-rich, maintaining habitable conditions despite the so-called faint young Sun problem (Haqq-Misra et al., 2008). Around 2.3 Gyr, the Earth underwent a great oxidation event (or multiple oxidation events) at the planetary scale, linked to the evolution of oxygenic photosynthesis by cyanobacteria and to geological processes permitting $O_2$ accumulation (Catling et al., 2001; Holland et al., 2002). This event involved the shift in redox conditions from $O_2$-free atmosphere to moderately oxidizing conditions typically containing 0.1% $O_2$ (Catling et al., 2005). This great oxidation appears so far decoupled from that of the Earth's mantle mentioned in Section 3.4, because almost 2 Ga separate both events. However, a possible tectonic regime transition from the Archean with microplates floating on a low viscosity mantle to the Proterozoic regime with deeper subduction and mantle mixing may have facilitated oxygenation of the upper mantle by subduction of iron oxides (Mikhail and Sverjensky, 2014). Initiation of deep subduction would also have allowed the ascent of oxidized lower mantle material towards the Earth's surface; both could have contributed to the GOE (Andrault et al., 2018). This appealing model (Andrault et al., 2018) however, implies that the oxidation state of volcanic rocks or of their mantle sources must have changed during or just before the end of the Archaean. So far, this long-standing hypothesis is not fully supported since available geochemical observations indicate that the oxidation states of mantle melts reaching the surface have remained constant from the early Archaean to present-day (Canil, 1997; Gaillard et al., 2015; Nicklas et al., 2018). This redox constancy in the oxidation state of the Earth's mantle remains however unexplained. It may simply reveal our inability to capture the signature in ancient rocks of any redox modifications or, that another process may have modified the oxidation state of volcanic gases without requiring a change of the mantle source. For example, Gaillard et al (2011) have shown that the conditions of volcanic degassing may trigger important changes in the oxidation state of volcanic gases while the magmatic source remains unchanged; this does not exclude a change in the oxidation state of the source, but it adds a degree of freedom in the processes linking the deep Earth to its surficial chemistry. Changes in the solid Earth that may have triggered the GOE could also include higher rates of serpentinization on more ultramafic seafloor and lower rates of ferrous iron delivery at mid-ocean ridges (Kasting, 2013a).

Earth redox conditions remained heterogeneous in space and time, with redox-stratified oceans and slight oxygenated atmosphere for more than a billion years before reaching modern conditions of fully oxic ocean and atmosphere only 800-600 Ma ago (Lyons et al., 2014).

Over longer, geological timescales, the biosphere on its own cannot change Earth's net global oxidation state because every biologically generated oxidant is accompanied by a mole-equivalent reductant. Carbon stored in biomass reservoirs, for example, is released back into the atmosphere-ocean system during e.g. organic decay or/and respiration. Instead, a net atmospheric redox shift requires that these redox products couple differentially to geologic fluxes, for example by increasing carbon sinks and burial of organics during tectonic events (Van Valen, 1971; Catling et al., 2001). The "carbon rain" ("marine snow") of organic material from the upper ocean layers down to the ocean surface can be subducted via burial into the Earth's mantle (a process that is likely to be much more efficient on a planet with plate tectonics). The removed organic material - which would otherwise have combined with oxygen during respiration - therefore implies a net excess of $O_2$(g).

Plate tectonics also has a strong influence on the continuous existence and diversity of volcanism. This is linked to subduction of hydrated material, which is released in the mantle wedge and triggers arc volcanism close to the subduction zone. Changes in the organic and inorganic components of the carbon cycle would have affected key gases in Earth's early atmosphere ($O_2$, $CO_2$, $CH_4$ and $N_2$), and are linked to the evolution of life.

The actual $O_2$, $CH_4$ and $N_2$ levels on Earth can be understood by considering their evolution in a global redox situation.

After the continents formed, the $CO_2$ cycle was influenced by both continental weathering (carbon-silicate cycle) and (to a smaller extent) seafloor weathering. On early Earth, seafloor weathering was likely the dominant weathering process (Kasting, 2013b; Coogan and Gillis, 2013; Coogan and Dosso, 2015; Krissansen-Totton and Catling, 2017).

Regarding the atmosphere, life interacts with biogeochemical systems in numerous ways, which we will only briefly describe here. In the carbon cycle (Falkowski et al., 2000) for example, phytoplankton absorb aqueous $CO_2$ in the upper ocean layers, which leads to a "biological pump" favoring absorption of $CO_2$ from the atmosphere into the ocean. Biomass and soil contain about three times the amount of carbon than what is present in the atmosphere, hence they influence carbon storage and atmospheric-ocean fluxes. Regarding the oxygen cycle (see e.g., Catling and Claire, 2005; Falkowski and Godfrey, 2008), the main atmospheric source (biological) is regulated by burial which controls an imbalance between primary production and respiration from the biosphere (as discussed above). Finally, the evolution of the nitrogen cycle may have had an important role in Earth habitability (Stüeken et al., 2016a) and is controlled by the evolution of Earth redox-state. $N_2$ is important in atmospheric pressure, itself controlling the stability of surface liquid water. Several nitrogen species (nitrous oxide $N_2O$, nitrogen dioxide $NO_2$, nitrogen gas N2, and ammonia NH3) are important greenhouse gases, or affect the production of ozone ($O_3$), also a greenhouse agent, or combine with organic compounds to produce aerosols hazes blocking solar insolation (Stüeken et al., 2016a). Microbes play a crucial role in fixing inert molecular nitrogen from the atmosphere into ammonium, which can subsequently be recycled back into the atmosphere as nitrous oxide (see e.g., Kasting, 2013b; Canfield et al., 2010; Yung and De More, 1999; Ward, 2012). Biomass emissions on Earth include e.g. oxygen, carbon dioxide, methane, nitrous oxide, chloromethane and isoprene, and sulfur-rich aerosols, which can affect atmospheric composition, climate and cloud cover, hence habitability. However, some of these gases are biologically produced only recently on the geological timescale, and only by complex eukaryotic algae. Before the GOE (~2.5 Ga), probably only methane was an important gaseous contribution of life to the atmosphere composition (Catling et al., 2001). With the appearance of cyanobacteria, the biological contribution of molecular oxygen by oxygenic photosynthesis, and molecular nitrogen and nitrous oxide by nitrification and denitrification increased in the atmosphere. Sulfur-rich aerosols (DMS: dimethyl sulfide, oxidized into sulfur compounds serving as nuclei for cloud formation) are produced by eukaryotic algae (haptophytes and some dinoflagellates) only since the Cenozoic (or the Neoproterozoic, see Feulner et al., 2015), but also by volcanic activity, and have an impact on climate, perhaps even triggering the Neoproterozoic snowball Earth although this remains speculative. The origin and evolution of these biogeochemical cycles on the early Earth has important implications for the characterization of plausible spectral biosignatures to detect in exoplanetary atmospheres (see below), and the possibility of complex life (active mobile organisms like metazoans) beyond Earth (Catling et al., 2005).

## 5.2. Other terrestrial planet atmospheres

Understanding our Solar System Planetary Atmospheres is a significant step forward for paving the way for future studies of atmospheres of Extrasolar Planets. Notably, Venus and Mars are natural comparative laboratories to investigate the diversity of atmospheric mass, composition and circulation regimes. In this context, comparative studies are useful to understand the evolution of climate of our Earth, both in the past and in the future.

Emissions of atmospheric greenhouse constituents such as methane could have played an important role in the early evolution of the Earth, by preventing it from becoming a frigid snowball (Pavlov et al., 2000; Kasting, 2013a). Serpentinization is one of the most efficient ways to create $CH_4$ abiotically (Etiope and Sherwood Lollar, 2013; Oze and Sharma, 2007) and it could have happened also on Early Mars with stronger volcanic activity, higher rates of large impact cratering, and liquid water temporarily flowing on the surface compared to present. This metamorphic process, by which low-silica mafic rocks are hydrothermally altered, produces serpentine and magnetite by storing water, and releasing $H_2$. The released $H_2$ reacting with oxidized carbon compounds, such as $CO_2$ and CO, can create $CH_4$ in reduced conditions through Fischer–Tropsch-type reactions (Etiope and Sherwood Lollar, 2013). The detection of methane in Mars atmosphere up to ~ 50 ppb locally (Mumma et al., 2009) indicates the presence of highly localized and sporadic $CH_4$ sources that could have been trapped in clathrates which are in stable form at depth (Chastain and Chevrier, 2007). The scientific objectives of the ESA's ExoMars programme include investigating Martian atmospheric trace gases and their sources. A particular emphasis is given to Methane and its origin. A potential biological origin would indicate the existence of long-extinct microbes or methane-producing organisms still active (Vago et al., 2017). An abiotic origin on the other hand, would have implications for the geochemical processes, atmospheric evolution and environmental conditions on early Mars (Chassefière et al., 2013; 2016) and Martian remanent magnetic field (Lasue et al., 2015).

## 5.3. Influence of life on the interior evolution of planets

Major shifts in Earth's evolution led to progressive adaptations of the biosphere. Conversely, effects of the emergence and evolution of life significantly affect the Earth's system (atmosphere, hydrosphere, geosphere, and mantle). The emergence of continents permitted more widespread use of solar energy. Land plants and benthic mats on exposed continents, but also mats and phytoplankton in oceans, use solar energy to convert $CO_2$ into biomass via photosynthesis, with $O_2$ as a by-product. This biomass is consumed by other organisms in soils and the oceans. Most terrestrial biomass is ultimately exported to the ocean as dissolved and particulate organic matter, and preserved in shales, unless oxidized or eroded (Wallmann and Aloisi, 2012). Tectonic activity creates relief, enhancing weathering that liberates carbon from organics or carbonates. Indeed, biologically enhanced weathering rates of silicate rock (Schwartzman and Volk, 1989; Drever, 1994; Hoffland et al., 2004) are likely crucial for the evolution of plate tectonics planets in various respects. Weathering is also an important component in the long-term silicate-carbonate cycle, which stabilizes Earth's climate. In this context, the biological enhancement of weathering rates has been argued to lengthen the lifespan of the biosphere (Lenton, 2002). In addition, the chemical dissolution of rock enhances the rate of surface erosion and thus the flux of sediments into subduction zones, including biological carbon. This establishes potential links to the deep interior (Rosing et al., 2006; Sleep et al., 2012; Höning et al., 2014). Water stably bound within the sediments contributes to the subduction rate of water into the Earth's mantle. The mantle responds by enhancing its rates of convection, water outgassing, and subduction. Water that is released in subduction zones at a depth of 100-200 km induces partial melting and therefore the production of new continental crust (Höning et al., 2014; Höning and Spohn, 2016). Importantly, this causes a positive feedback cycle, since more continental crust is subjected to weathering. Part of the geological carbon cycle is controlled by life and is also important to consider in Earth's habitability. On a long time-scale, the most relevant fluxes in recycling organic carbon are the weathering of fossil organic carbon on land and burial of organic matter, carbonates and methane in marine sediments (Wallmann and Aloisi, 2012). Metamorphism of subducted carbonate rocks and fossil organic carbon, and mantle degassing at subduction zones, hot spots, and volcanoes are important sources of $CO_2$, in the atmosphere, while life and weathering as important sinks, as explained above. As the biosphere enhances weathering, it can be argued to play an important role in preserving the present-day Earth state with large emerged continents. An Earth-like planet without biosphere could evolve into a water-world scenario with no or few emerged continents as suggested by Höning and Spohn (2016). Mechanisms related to the presence of life on the surface of a planet like Earth are shown in Figure 5.

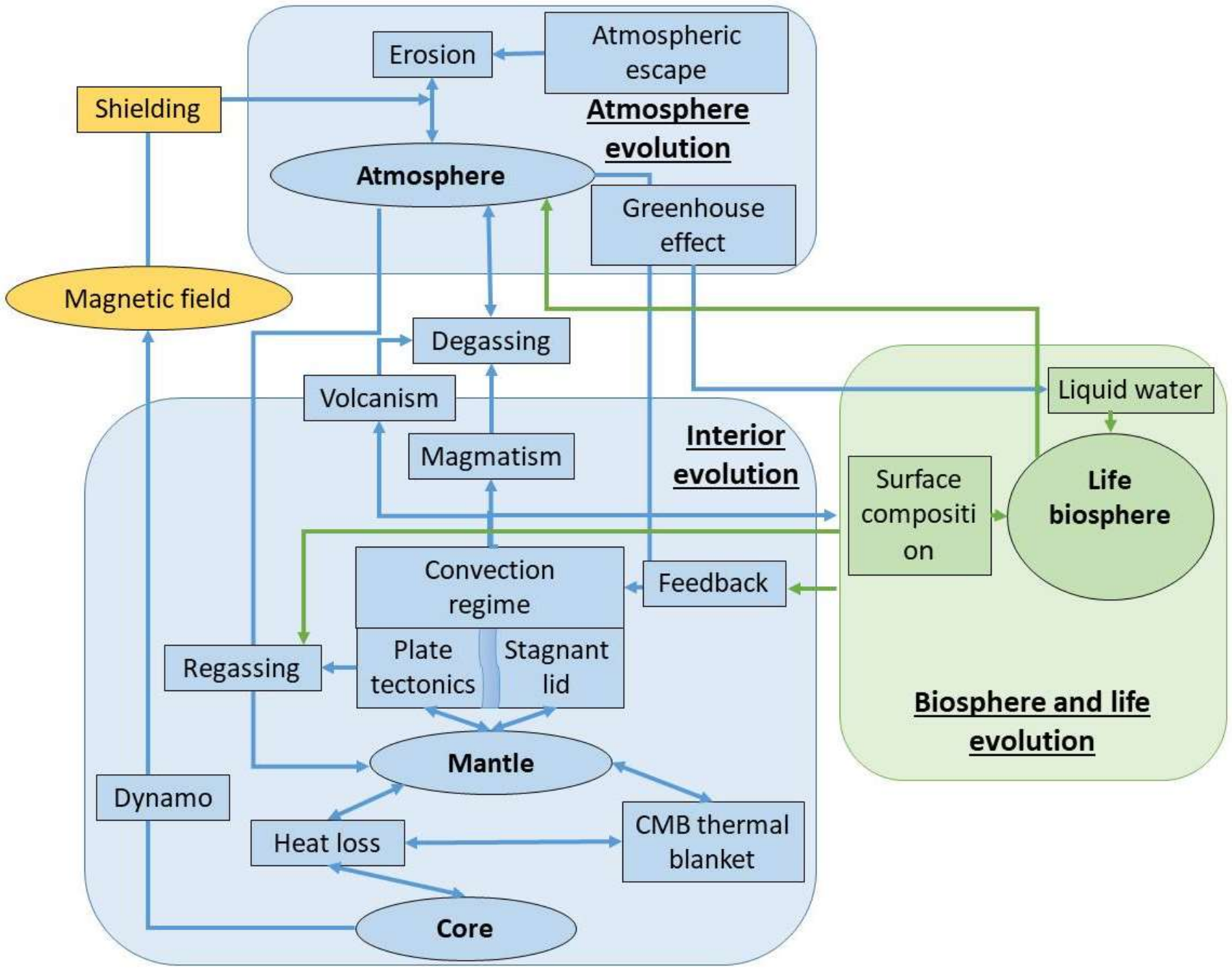

*Figure 5: Habitability and heat/interior/atmosphere evolution, and the potential role of life.*

## 5.4. Combined effects on atmosphere and interior evolution

Having considered various factors individually, we graphically link their combined effects in Figure 6.

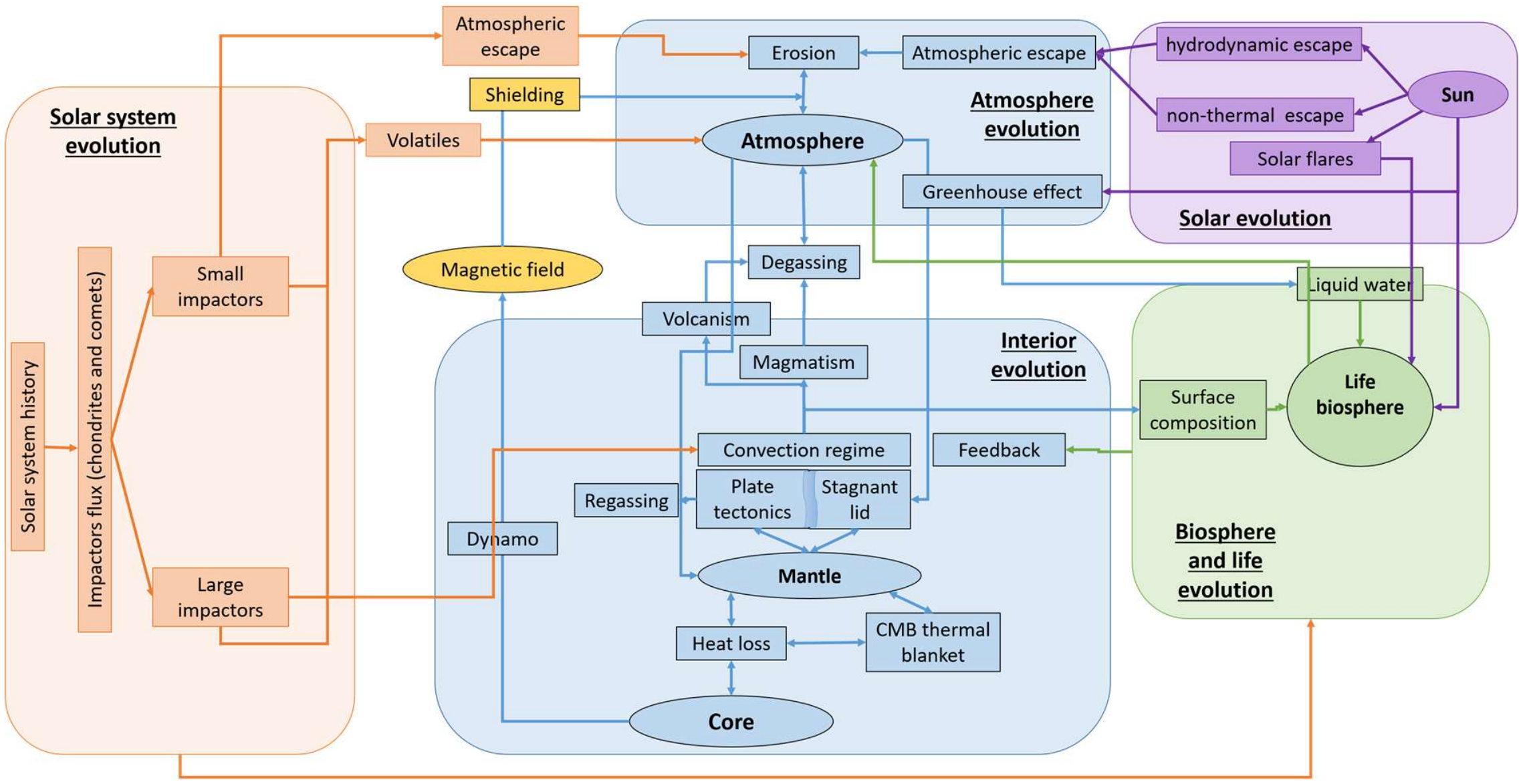

*Figure 6: Habitability and heat/interior/atmosphere evolution, and the role of impacts on life.*

Figure 6 illustrates the potential complexity of processes that could influence planetary habitability. It is clear that an interdisciplinary approach is required in order to elucidate the potentially subtle feedbacks that can operate.

# 6. Identification of preserved life traces in the context of the interaction of life with planetary evolution

## 6.1. Life's Origin and Environmental Tolerances

The origin of life remains one of the most challenging questions and although it will never be known for certain, we can still propose plausible scenarios. Whether life appeared in a hot or lukewarm, deep or surface environment, life is a natural phenomenon, and resulted from complexification of organic matter (Reisse, 2011). Elements (C,H, N, O, P, S, metals) and abiotic organic molecules are common in the universe and form through various processes in space and on Earth, including some of the molecular building blocks of life (amino acids, bases, sugars). The origin of water on Earth is reasonably well established (see above), and comes mainly from carbonaceous chondrites and degassing. Earth has been habitable (with liquid water and some crust) since the Hadean, as evidenced by geochemical and mineralogical analyses of 4.3 Ga zircons (review in Martin et al., 2006; Whitehouse et al., 2017), and thus life could have originated soon after the cooling and differentiation of our planet. Various scenarios propose that life originated with the appearance of a (lipidic or mineral) membrane first, of nucleic acids first (genetic code, proteins or RNA before DNA), or metabolism first. The plausibility of these scenarios can be tested in light of the geological conditions prevailing in the early Earth, although we have little information for the Hadean, from which rocks are not preserved, but only minerals (zircons). For a review, see e.g., Hazen (2012); Bada and Lazcano (2002); Szostak (2017).

Life on Earth has adapted to a wide range of physico-chemical conditions (e.g. Javaux and Dehant, 2010). Organisms called "extremophiles" not only tolerates but also thrives in extreme values of these parameters (temperature, salinity, oxygen concentration, pressure, vacuum, pH, radiation, metal concentrations …), offering a perspective on the limit of life on Earth. Some of the metabolisms allowing adaptation to these ecological niches appeared early in Earth history, but many are derived, e.g. appeared as secondary adaptations and cannot be used as analogs of early life or extraterrestrial life. Another point to consider is how these extremophiles benefit from nutrients originating from less extreme environments transported by wind or water. Carbon-based life needs, at the minimum, liquid water in contact with minerals, where nutrients (elements) are concentrated. To replenish the sources of nutrients, recycling is necessary, and thus perhaps plate tectonics or some sort of geological activity. Some forms of life, but not all, also need light. A much wider range of environmental parameters certainly exists on planetary bodies within and beyond our Solar System and the question arises which set of environmental parameters would still allow the origin and persistence of life.

## 6.2. Early Earth

The search for the existence of life on the early Earth or beyond Earth requires the characterization of biosignatures, or "indices" or "traces" of life. These traditionally include fossil chemicals produced only by biological activity, isotopic fractionations of elements indicative of biological cycling, biosedimentary structures induced by microbial mats such as stromatolites, and microstructures interpreted as morphological fossils. However, these traces can also be produced by abiotic processes or later contamination, in some cases, leaving a controversy surrounding the earliest record of life on Earth. Only the combined characterization of possible and multiple traces of life and their geological context, from the macro- to the micro- and nanoscales, can increase confidence in the fossil record and help discard abiotic explanations.

Geobiological studies can improve our understanding of preservational environments and taphonomic processes (how organisms decay and become fossilized), abiotic processes and products, and help us to develop a multidisciplinary approach to establish the biogenicity (biological origin), endogenicity (the fact that the microfossil or other trace of life is not a contamination in the rock), and syngenicity (the fact that the trace of life has the same age of the host rock) of these *in situ* biosignatures or the possible biogenicity of atmospheric signatures (e.g., Brasier et al., 2006; Javaux and Lepot, 2017; van Zuilen et al., 2002; Javaux, 2019). Combining minimum ages of fossil biosignatures with molecular phylogeny ("tree of life") permits to produce molecular clocks, that provide dating of branching events and important biological innovations, and allow predictions for the evolution of former and later clades or metabolisms. The analysis of these fossils and other traces of life has allowed the reconstruction of the co-evolution of Earth and life. This suggests that the Earth was inhabited early in its history, since at least 3.5 Ga (Allwood et al., 2009). Among the oldest unambiguous traces of life preserved on Earth, stromatolites, isotopic fractionation of C, S, N, and microfossils indicate the presence of a microbial biosphere at least since 3.45 Ga (review in Knoll et al., 2016; Javaux, 2019). Claims for older traces range up to 4.3 Gyr but are ambiguous and could be explained by alternative abiotic processes (e.g., McCollom et al., 2001; van Zuilen et al., 2002; García-Ruiz et al., 2009; Grosch et al., 2017; Wacey et al., 2018). Such putative traces include the presence of organics with carbon isotope fractionation in zircons (Bell et al., 2017) or metamorphic rocks (Tashiro et al., 2017), folded structures interpreted as stromatolites in deformed metamorphic rocks (Nutman et al., 2016), or mineral tubes interpreted as microfossils (Dodd et al., 2017), or traces interpreted as endolithic burrows in pillow laves (Furnes et al., 2004). Within possible windows of preservation relevant both for early Earth and early Mars (see below), fine-grained (shales and siltstones) or coarse-grained (sandstones) siliciclastic rocks preserve very well fossil organic remains through the geological record, including microscopic cells and microbial mats as old as 3.2 Ga (Javaux et al., 2010; Homann et al., 2015; Noffke et al., 2006), photosynthetic pigments as old as 1.6 Ga (Brocks et al., 2005; Gueneli et al., 2018) and eukaryotic steranes as old as 800 Ma (Brocks et al., 2016; Love et al., 2009. Probable abiotic organics have recently been reported from 3 Ga Mars samples (Eigenbrode et al., 2018). Carbonates, chert (Sugitani et al., 2010; Westall et al., 2011) and evaporites may also preserve traces of life, but chert are formed by hydrothermal fluids, that can also lead to the formation of abiotic organics migrating around minerals and creating abiotic pseudofossils (e.g., Van Zuilen et al., 2002;

Knoll, 2016) or preserving volcanic glasses resembling tubular or lenticular microfossils (Wacey et al., 2018; Hickmann-Lewis et al., 2018), and carbonates can host folded structures or mineral precipitates that can be confused with stromatolites. Evaporites are very soluble, but can be sometimes preserved in the geological record (e.g. Allwood et al., 2013). Some of these lithologies are common on the early Earth, and may be relevant to Mars (Westall, 2008; Vago et al., 2017; McMahon et al., 2018).

The preservation of molecules such as pigments, signature of photosynthetic metabolisms, can be used to track life's traces on the early Earth, and as analog studies for the detection of complex biological molecules in early Mars sedimentary rocks (e.g. Storme et al., 2015). Modern cyanobacterial mats growing in Antarctica provide an opportunity to study microbial biosignature preservation in clastic environment exposed to high UV radiation, and undisturbed by metazoans (Lepot et al., 2004).

As discussed above, cyanobacteria are crucial microorganisms because they "invented" oxygenic photosynthesis, changing forever the chemistry of early Earth oceans and atmosphere and creating new ecological niches where mobile complex life diversified. Determining new biosignatures of cyanobacteria might help to constrain the timing of oxygenic photosynthesis and to pinpoint the origin of the chloroplast, resulting from an important symbiotic event that led to a large diversification of the photosynthetic eukaryotes (Demoulin et al., 2019). The characterization of biosignatures from cyanobacterial mats growing in high UV Antarctic lacustrine habitat showed that their cells and filamentous sheaths with diagnostic pigments (scytonemin) can be preserved in 4000 years old siliciclastic sediments (Lepot et al., 2014). The preservation of these organic remains occurred through the precipitation of nanominerals (aragonite and clays).

Complex organic molecules and simple and complex organic morphologies, and also abiotic organics from meteorites or geological processes, can thus be preserved in unaltered mudstones or shales and can be searched for in ancient Earth rocks or in extraterrestrial rocks with similar lithologies (e.g., Javaux et al., 2010; McMahon et al., 2018; Eigenbrode et al., 2018).

## 6.3. Mars

Searching for life (past and present) on Mars is one of the main motivators for exploring the planet.
The preservation of organic remains in clay minerals in early Earth samples and possibly analogue to those deposited during the Noachian on Mars permit the identification of strategies for detection of biosignatures on Mars by the future Martian rovers. Indeed, the rovers are using a combination of macro-to microscopic tools to characterize the geological context and possible habitability of local martian areas and lithologies, such as panoramic cameras, close-up imager (like the future CLUPI-a ExoMars instrument, Josset et al., 2017, microscopes, IR and RAMAN microspectrometers to identify minerals and possible (abiotic or biotic) organics, and mass spectrometers measuring C isotopes). Because of the destructive effects of the strong radiation environments at the surface of Mars, organic biosignatures need to be looked for under sediment or rock surfaces, as proposed for the ExoMars rover equipped with a drill. Recently, organics were detected in 3 Ga mudstones of the Martian Gale crater by the NASA rover curiosity (Eigenbrode et al., 2018). As discussed above, the mere presence of organics is not at all indicative of biological origin, but this discovery shows that they can be preserved in ancient Martian sediments. Collection of well-characterized fossil-rich and fossil-barren rocks can serve as reference to test instruments and landing and sampling strategies for future missions.
Westall et al. (2015) provided a recent review which summarizes how, where and what to search for. The future ExoMars 2020 mission includes a rover that will search for life signs.

The Martian surface is subject to intense UV irradiation as well as and features a highly oxidizing and acidic soil. How organism would respond to such conditions can be measured in laboratory experiments designed to simulate planetary (and asteroid surface) conditions. The Trace Gas Orbiter (Korablev et al., 2014) (part of the ongoing ExoMars mission) will deliver groundbreaking data on the evolution, composition and climate of Mars' atmosphere.

## 6.4. Earth-like Exoplanets

The process of defining a scientific roadmap towards detecting biosignatures on Earth-like exoplanets has considerably expanded and matured in recent years. The first step is to compile a suitable target list. This involves identification of an appropriate sample of rocky planets lying in the Habitable Zone of their central stars - which is a central task of the PLATO 2.0 mission (Rauer et al., 2014). The so-called eta-Earth ($\eta_{Earth}$) parameter describes the percentage occurrence rate of Earth-like planets around main sequence stars. For example, 3.2 ± 1.6 % of F,G,K stars are estimated to have planets with up to two Earth radii with orbits from (100-200) days (Petigura et al., 2013), and an even higher percentage of stars would be expected to have planets in their entire habitable zone.

The second step is to perform the measurement itself. For this, several remote methods have been proposed. For example, atmospheric spectrophotometry has been examined by numerous studies as a possible technique to measure spectral features of potential gas-phase biosignatures such as oxygen ($O_2$), ozone ($O_3$) and nitrous oxide ($N_2O$) (see e.g., Des Marais et al., 2002; Segura et al., 2003; Tinetti et al., 2006; Kaltenegger et al., 2010; Rauer et al., 2011; Grenfell et al., 2013; Meadows, 2017; Gillon et al., 2016; 2017 to name but a few). However, these biosignatures result from complex metabolisms that evolve relatively late on Earth (around 2.5 Ga), hence are less probable to appear also beyond Earth. The plausibility of Earth-like spectral biosignatures in exoplanet atmospheres should be assessed in light of the origin and evolution of these biogeochemical cycles on the early Earth (see Section 5.1). Surface spectroscopy has been proposed to search for e.g. leaf reflectance (the "red edge", see e.g. Seager et al., 2005), to look for biopigments (Schwieterman et al., 2015) or even to search for microorganisms (Hegde et al., 2014). Spectropolarimetry has been suggested as a technique to detect photosynthetic biopigments (see e.g. Berdyugina et al., 2016). An excellent summary of recent advances in exoplanetary biosignature science can be found in a series of review papers from the NASA Nexus for Exoplanet System Science (NExSS) and Astrobiology Program, namely: Schwieterman et al. (2018) who provide an all-round review of the topic; Meadows et al. (2018) who focus on oxygen; Catling et al. (2018) who discuss a Bayesian approach; Walker et al. (2018) who propose future methods for biosignature assessment and Fujii et al. (2018) who summarize future relevant missions. A review of atmospheric biosignatures with a focus on photochemical responses in an exoplanetary context is provided by Grenfell (2017). In addition to considering spectral bands of individual gas-phase species, one can alternatively search for two (or more) species present simultaneously with relative abundances that can only be accounted for by invoking the existence of life. Examples are e.g. the species pair ($CH_4$ and $O_2$). Without life on Earth to re-supply them, these species would react and be removed to form carbon dioxide and water, see e.g. Krissansen-Totton et al., 2018). Krissansen-Tottet et al. (2018) and Lammer et al. (2018) discuss nitrogen and oxygen cycles in a biosignature context. Kasting et al. (1993) noted that nitrate removed to an abiotic ocean would be reduced to either $N_2$ or $NH_3$ during passage through the mid-ocean ridge hydrothermal vents, and $NH_3$ would be photochemically converted back to $N_2$. Thus, the presence of $N_2$ in a planet's atmosphere does not require the existence of biological denitrification.

The third step is to retrieve atmospheric state variables (e.g. p, T, composition) by performing a backwards calculation starting with the observed spectra (see e.g., von Paris et al., 2013; Barstow et al., 2016) combined with an estimate for noise sources (e.g. photon noise, instrument noise etc.). Chemical species with strong, unique spectral features i.e. without overlapping bands are therefore favored during this process to avoid convergence issues and numerical degeneracies (see e.g. Benneke and Seager, 2012).

The fourth step is to interpret and assess biosignature candidate signals. This involves identifying and discounting so-called biosignature false positives (i.e. dead planets which nevertheless mimic the presence of life) and false negatives (i.e. planets with life which is not detected) (see e.g., Selsis et al., 2002; Schwieterman et al., 2016). Useful in this context are estimates (e.g. from model studies) of the atmospheric climate and composition of a so-called "dead Earth" i.e. a planet with the same properties as Earth except where life did not develop (see e.g., Yung and De More, 1999; Margulis and Lovelock, 1974) in order to have a benchmark against which to compare. There exists additional data which is useful to have when assessing potential biosignatures, including the stellar class, metallicity, the age of the system, the planetary (mass, interior), orbit and atmospheric properties (mass, composition, temperature, albedo, clouds etc.) hence climate and habitability. Many of these issues have been discussed e.g. in the special issue on "Planetary evolution and life" (Spohn, 2014 and references therein).

There exists a range of (exo)planetary parameters proposed to be related to the evolution of life for which observational techniques have been suggested. Some of these however are still in an early, discussion stage. For example, the bulk atmospheric composition could be derived from measuring the so-called Rayleigh slope (see e.g. Kreidberg et al., 2014) arising due to atmospheric scattering in the transmission spectrum in the UV/visible. The absorption spectra of key molecular absorbers such as $CO_2$, $H_2O$ can be derived during the primary transit (i.e. where the planet passes in front of the star) and from this, one can estimate species' abundance profiles. From the secondary transit (where the planet disappears behind the stellar disk), one can constrain e.g. the planetary temperature and climate-related profiles. From reflection spectra and/or planetary phase curves (see e.g. Selsis et al., 2011) one can constrain clouds, albedo and atmospheric circulation. Some works (e.g. Vidotto et al., 2010) suggested that the exoplanetary bow shock could be constrained from the early UV ingress of the planetary transit. Planetary reflection spectra could in principle deliver information about surface mineralogy and help distinguish between ultramafic, hydrated or/and water ice surfaces (see e.g., Mukhin et al., 1989; Hu et al., 2012) or even between ocean, soil, snow and vegetation based on isotropic scattering model (Fujii et al., 2010). Robinson et al. (2014) reported a detection of Earth's ocean glint from the Lunar CRater Observation and Sensing Satellite (LCROSS) mission and discussed the consequences for Earth-like exoplanets.

Biosignature detections for next generation missions such as the James Webb Space Telescope (JWST) (see e.g. Barstow et al., 2016) and the European-Extremely Large Telescope (E-ELT) are estimated to be very challenging (e.g. Rodeler and López-Morales, 2014) but could be possible if suitable targets are found. Somewhat farther into the future are missions discussing rocky, Earth-like planets and their potential biosignatures, including the Large UV/Optical/IR Surveyor mission (LUVOIR, France, 2016), the Habitable Exoplanet Imaging Mission and the Origins Space Telescope. Further details of these missions are summarized in Fujii et al. (2018).

## 7. Conclusion on habitability and planet formation in a broader context

The search for and characterization of Earth-like planets in the habitable zone of stars has become a central focus of research in astronomy, astrobiology and the planetary sciences although life may also have formed in water-rich planetary objects outside of habitable zones. Understanding whether a planet could be potentially habitable – and understanding the habitability of Earth – requires a deeper knowledge of the geophysical processes driving the key elements for habitability as developed in the present paper. To gain a better understanding of these processes, the evolution of Earth is often taken as a reference case for the interaction of atmosphere, geology and biological processes. Such processes are expected to take place on terrestrial exoplanets also, but are much harder to constrain without *in situ* information. Furthermore, terrestrial planets which orbit different stars of different spectral classes, in particular M-dwarf stars, or/and young planetary systems, may experience much harsher space weather conditions that could affect habitability as well as the presence of biosignatures.

Terrestrial exoplanets are currently starting to be observed around different stars. Terrestrial planets can have deep or shallow oceans; they have been suggested to possess hydrogen-rich atmospheres or could lose their entire atmospheric inventories due to loss processes such as escape. Spectroscopic measurements during planetary transits and occultations give insights into the composition of exoplanetary atmospheres and albedos, which may help in hypothesizing the presence of oceans and lands, and might give some hints on the interaction between surface and interior. These measurements however must be further studied in light of the geophysical context, which affects atmospheric evolution as discussed in the previous section.

The recent discovery of seven roughly Earth-sized planets orbiting the ultra-cool, low-mass star TRAPPIST-1 (Gillon et al., 2016; 2017) has vaulted this system to the forefront of exoplanetary characterization. The planets orbit the star with semi-major axes < 0.1 AU, and orbital periods of a few Earth days. Given their proximity to the star, and the star's low mass and low luminosity, each TRAPPIST-1 planet has a moderate effective temperature (from ~160 to 400 K, Gillon et al., 2017). Depending on their atmospheres and possible greenhouse effects, the surfaces can either be solid composed of water ice and/or rock, or could support liquid water (Gillon et al., 2017; Checlair et al., 2017; Wolf, 2017; Turbet et al., 2018). The planets' orbits are in a near mean motion resonance, which maintains their eccentricities, raising tidal forces in the bodies that heat their interiors by tidal dissipation. Tidal heating may be an important energy source for the innermost planets that can significantly increase their temperatures (Luger et al., 2017, Gillon et al., 2017, Barr et al., 2018). Interior models can help in estimating the possible structures of the planets and give some insights into their possible habitability (see Barr et al., 2018). Similar planetary systems to TRAPPIST-1 are expected to be discovered in the next few decades, because M dwarfs are the most numerous stars in the Galaxy and the detectability of their close-in planets makes them favorable targets for investigating possible life-harboring worlds, although the effect of higher radiation rates and tidal locking on habitability need to be assessed. Due to the low luminosity of the star, the habitable zone is located in a very close distance around the star. For this reason, the planet/star size ratio for planets in the habitable zone is large, making their detection easier. Current and upcoming missions will find new exoplanetary systems, and their detailed modelling will be an important task to reveal their astrobiological potential.

## Acknowledgements

This paper was triggered by presentations and discussions that were held during the conference "Geoscience for understanding habitability in the solar system and beyond" held in the Azores from the 25th-29th September 2017. The workshop (and the work behind the scientific presentations) was supported by:

- European COST (Cooperation in Science and Technology) Action TD1308 "ORIGINS" (Origins and evolution of life on Earth and in the Universe),

- EGU (European Geophysical Union) Galileo conferences,
- EuroPlaNet (European Planetology Network) 2020 RI (Research Infrastructure) (EPN2020-RI),
- German TRR 170 (TransRegio collaborative research) network,
- GINOP-2.3.2-15-2016-00003,
- Hungarian National Research, Development and Innovation Office (NKFIH) grants K119993, K-115709,
- The Austrian Science Fund (FWF) NFN project S11601-N16 “Pathways to Habitability: From Disks to Active Stars, Planets and Life” (related sub-projects S11604-N16, S11606-N16 and S11607-N16),
- Planet TOPERS (Planets: Tracing the Transfer, Origin, Preservation, and Evolution of their ReservoirS) Belgian IAP (Inter-university Attraction Pole) PAI-IAP P7/15,
- EU FP7-ERC Starting Grant ELiTE 308074: Early life Traces, Evolution, and Implications for astrobiology,
- FNRS FRFC T.0029.13 ExtraOrDynHa,
- ET-HOME (**E**volution and **T**racers of the **H**abitability **o**f **M**ars and **E**arth) Belgian Excellence of Science –EoS- program, EOS 30442502.

They are very much acknowledged, as well as anonymous reviewers for their helpful reviews.